\providecommand{\linenumbers}{}
\documentclass{aa}  
\usepackage{graphicx}
\usepackage{txfonts}
\usepackage{subcaption}
\usepackage{lscape}
\usepackage{placeins}
\usepackage{xcolor,ulem}
\usepackage{amsmath,amssymb}
\usepackage[utf8]{inputenc}
\usepackage[english]{babel}
\usepackage[breaklinks, colorlinks, citecolor=blue]{hyperref}

\begin{document}

\title{Constraints on Hadronic Emission from Microquasars Detected by LHAASO}

\author{
V. Vecchiotti\inst{1,2}\thanks{Corresponding author: vittoria.vecchiotti@inaf.it} 
\and E. Amato\inst{1} 
\and G. Giacinti\inst{3} 
\and G. Morlino\inst{1} 
\and G. Peron\inst{1}
}

\institute{
INAF Osservatorio Astrofisico di Arcetri, Largo Enrico Fermi 5, 50125 Firenze, Italy
\and
Tsung-Dao Lee Institute, Shanghai Jiao Tong University, Shanghai 201210, P. R. China
\and
Universit\'e Paris Cit\'e, CNRS, AstroParticule et Cosmologie, F-75013 Paris, France
}

\date{Received June 19, 2026; accepted --}

\abstract
{Recently, the Large High-Altitude Air Shower Observatory (LHAASO) collaboration reported ultra-high-energy (UHE, $>$100 TeV) gamma rays from six microquasars.
For five of these sources, the emission extends beyond $100$ TeV, making microquasars promising candidates for Galactic PeVatrons.}
{We investigate whether gamma-rays around $100$ TeV originate from hadronic interactions of accelerated cosmic rays with the ambient medium, and we estimate the contribution of these sources to the measured cosmic-ray proton spectrum around the knee.
We also derive upper limits on the contribution from the six additional microquasars in the LHAASO field of view, from which no UHE emission was detected.}
{We consider a cosmic-ray transport scenario in which particle propagation in the source vicinity is diffusion-dominated, with a diffusion coefficient largely suppressed with respect to the average Galactic value, while, at large distances from the source, the diffusion coefficient matches the Galactic value.
In addition, we assume continuous injection over timescales of $t_{\rm age}=0.1-1$ Myrs.
}
{Using available measurements of the gas density in the emission region, we find that hadronic interactions alone cannot fully account for the observed emission for any of the detected sources. 
However, in the case of GRS~1915+105 and MAXI~J1820+070, the hadronic scenario may still be valid when considering acceleration efficiency higher than 10\%.
We then derive upper limits on the hadronic contribution to the observed gamma-ray flux.
We estimate that the detected sources contribute at most $\sim4\%$ of the Galactic cosmic-ray spectrum at $1$ PeV for an injection timescales of $0.1$ Myrs and a cosmic-ray acceleration efficiency of $10 \%$.
When adopting the maximum acceleration efficiency allowed by the gamma-ray observations, instead, the contribution of these microquasars to the Galactic cosmic-ray spectrum at PeV energies rises to $37\%$.
%
Longer injection timescales ($\sim 1$ Myrs) lead to contributions exceeding the observational constraints.
For sources not detected at UHE, we obtained a maximum contribution of $\sim 17\%$, achieved assuming continuous injection over $1$ Myr.
}
{}


\maketitle

\section{Introduction}
\label{sec:intro}

X-ray binaries are binary systems in which a compact object, either a neutron star or a black hole, accretes matter from a companion star.
These systems can be in a quiescent state, believed to be associated with a low accretion rate, or undergo outbursts, during which the luminosity largely increases, sometimes exceeding the Eddington limit.
During outbursts, relativistic jets are launched and the system is in the so-called hard X-ray state, when the emission is dominated by the corona. The jets are quenched, instead, in the soft state, when the emission is dominated by the accretion disk \cite[see, e.g.,][]{Kalemci:2022fqq}. Systems hosting such jets are commonly referred to as microquasars.

Microquasars have long been observed in radio and X-rays and have been considered promising particle accelerators, potentially contributing to the Galactic cosmic-ray (CR) spectrum, e.g., at a few GeV \citep{Heinz:2002qj}.
Early hints of UHE emission date back to the 1980s, when a $4.4~\sigma$ signal of orbitally modulated PeV photons was reported from Cygnus X-3 \citep{1983CygnusX3}, although this result was not confirmed by later observations.

These systems have also been extensively investigated as potential neutrino sources, where high-energy neutrinos can be produced via $p$--$p$ or $p$--$\gamma$ interactions of accelerated CRs, either with the companion star \citep{1985NaturecygnusX3}, within the inner jet \citep{2010MNRAS.407.2468Z}, or through interactions between the jet and the stellar wind in case of high-mass X-ray binaries \citep{Romero:2003td,Romero:2005em}.

A major observational breakthrough came with the detection of TeV gamma rays from LS 5039 by HESS \citep{Aharonian:2005cx} demonstrating that microquasars are indeed efficient sites of particle acceleration.
Since then, several leptonic and hadronic models have been proposed to explain the broadband emission of systems such as Cygnus X-1 up to GeV energies \citep{Zhang:2014qda,Pepe:2015yxa}.
%
More recently, the HAWC collaboration reported the detection of TeV gamma-ray emission from SS~433 \citep{HAWC:2018gwz}, later confirmed by HESS \citep{HESS:2024rlh}. 
These observations are broadly consistent with inverse Compton emission from relativistic electrons in the jets.
Despite the success of leptonic scenarios, an hadronic interpretation of these sources remains viable.
In this case neutrino production is expected either in the jet \citep{Papavasileiou:2020lgf}, in the corona \citep{Fang_2024} or via jet-wind interaction \citep{Koljonen:2023xfn}, making these sources primary targets for multi-messenger studies \citep{Carpio:2025arz, Paul:2025tfd}.


%
Detailed modeling has shown that in principle, efficient diffusive shock acceleration, close to the Bohm limit, can reproduce the multi-wavelength emission of SS~433 \citep{Sudoh:2019jup}. These authors also discussed how acceleration of protons, along with electrons, would guarantee proton energies beyond PeV.
%

Evidence supporting hadronic processes as the origin of microquasar gamma-ray emission has emerged from the detection of multi-GeV radiation from GRS~1915+105 in Fermi-LAT data: here the emission appears consistent with interactions between relativistic protons accelerated in the jet and the surrounding medium \citep{Marti-Devesa:2024otf}.

Finally, the most recent observational breakthrough is the discovery by the LHAASO collaboration of UHE emission from 6 out of 12 known microquasars in the experiment field of view \citep{LHAASOmicroquasars,LHAASO:2025ysm}.

%
These findings establish microquasars as a compelling new class of Galactic PeVatrons.
In particular, the LHAASO collaboration showed that the UHE morphology of SS~433 appears correlated with the location of a nearby molecular cloud, hinting at a possible hadronic origin of the emission.
In this context, \cite{Abaroa:2025ege} proposed that microquasar remnants (sources that have recently switched off) could contribute significantly to the population of unidentified LHAASO sources \citep{LHAASO:2023rpg}, as CRs injected during their active phase may remain confined within their cocoons and interact with dense ambient gas.

Following these discoveries, microquasars have been increasingly investigated as potential contributors to the CR spectrum around the knee.
Various acceleration sites and acceleration processes have been shown to be capable of reaching PeV energies within microquasar environments.
%

For instance, acceleration may occur in trans-relativistic jets or winds with kinetic luminosities exceeding $10^{39} \, \rm erg \, s^{-1}$ \citep{Wang:2025yqy} , or in large-scale wind bubbles produced by super-Eddington accretion in X-ray binaries \citep{Peretti:2024ecg}.
Recollimation shocks forming in jets propagating through dense environments have also been proposed as efficient particle acceleration sites, capable of accelerating particles up to PeV energies and producing hadronic signatures through interactions in the surrounding cocoon \citep{Peretti:2026wqj}. 

Among the possible acceleration mechanisms, diffusive-shock acceleration can operate at different types of shocks, including wind termination shocks \citep{Peretti:2024ecg} and jet recollimation shocks \citep{Peretti:2026wqj}.
Another possibility is shear acceleration in jet-cocoon structures, which has been proposed as a mechanism capable of reaccelerating sub-TeV Galactic CRs and producing features consistent with the observed CR knee \citep{Zhang:2025tew}.

Despite these promising scenarios, the overall contribution of microquasars to the CR spectrum remains uncertain.
\cite{Peretti:2024ecg} found that a significant contribution to the CR knee would require the existence of a nearby, yet undetected, microquasar, while \cite{Kaci:2025gyb} argued that a relatively small population of active, powerful microquasars could already account for the observed proton flux around the knee.
Furthermore, recent work has shown that sources such as Cygnus X-3 may accelerate CRs beyond PeV energies, producing high-energy photons and neutrinos via $p$--$\gamma$ interactions and interactions with stellar winds \citep{Kachelriess:2025hkj}, in agreement with recent LHAASO observations \citep{LHAASO:2025ysm}.
At the same time, not all microquasars provide equally strong indications of hadronic processes.
%
For example, observations of V4641~Sgr from LHAASO \citep{LHAASOmicroquasars}, HAWC \citep{Alfaro:2024cjd} and HESS \citep{HESS:2025uyo} appear consistent with a predominantly leptonic origin due to the lack of sufficient target material in the surrounding environment \citep{HESS:2025uyo}.
This interpretation is further supported by subsequent studies, in which leptonic  \citep{Wan:2025eoh} or lepto-hadronic scenarios \citep{Kleimenov:2025bui} are preferred over a purely hadronic one.


This work aims to assess whether, or in which cases, the gamma-ray emission of the 6 microquasars detected by LHAASO can be interpreted as purely hadronic, irrespective of the acceleration mechanism.
We consider a transport scenario in which diffusion is suppressed in the vicinity of the source and quantify what fraction of the gamma-ray emission of each source can originate from $p-p$ interactions. We then estimate the contribution of these sources to the CR proton spectrum at PeV energies, assuming that particle injection is continuous over the source lifetime.

%
%

%
We first compute the grammage accumulated around microquasars and compare it with the values inferred from boron-to-carbon measurements \citep[e.g.,][]{Aloisio:2015,Blasi:2017,Bresci2019,Evoli:2019wwu, Ambrosone:2025wxc}, following the approach of \cite{Blasi:2025yjl}.
In particular, we adopt the value reported by \cite{Evoli:2019wwu}.

We then estimate the amount of target material required to reproduce the gamma-ray flux observed by LHAASO under the two considered transport scenarios, and compare these requirements with available gas measurements around these sources.

Finally, for the five sources detected above 100 TeV, we estimate the expected contribution to the cosmic ray proton knee, while for the remaining seven sources non-detected by LHAASO above 100 TeV, we derive an upper limit after constraining their luminosity.

The paper is structured as follows. In Section \ref{sec: LHAASO's microquasars}, we summarize the properties of the investigated microquasars. In Section \ref{sec: Method}, we describe the transport model (Section \ref{sec: Transport}), the   calculation of the target density required to produce the observed gamma rays via p-p interactions (Section \ref{sec: Grammage}), and the measured gas density (Section \ref{sec: measured gas}). Our results are presented in Section \ref{sec: Results} and discussed in Section \ref{sec: Discussion}. Finally, we summarize our conclusions in Section \ref{sec: Conclusion}.

\section{Microquasar sample}
\label{sec: LHAASO's microquasars}

In Table \ref{table sources}, we report the properties of the microquasars considered by \cite{LHAASOmicroquasars, LHAASO:2025ysm} located in the Northern sky and accessible to the LHAASO observatory, including both detected and undetected ones. The values reported in the table are the ones we use in the following calculations.

%
The mass of the compact object, $M_{\rm BH}$, and the distance from Earth, $d$, are taken from the black hole transient catalog BlackCAT \citep{BlackCAT}, except for SS~433 and Cygnus X-3, which are not in the catalog.

The spatial extension of the gamma-ray emission, $R_{\rm s}$, is derived from the angular extension reported in Table 1 of \cite{LHAASOmicroquasars} and in \cite{LHAASO:2025ysm} for Cygnus X-3.
For SS~433c, V4641~Sgr, and GRS~1915+105, $R_{\rm s}$ is defined as the 39\% containment radius, while for MAXI J1820+070 is the one-tailed  95\% confidence upper limit.
For Cygnus X-3, $R_{\rm s}$ corresponds to the physical size of the gamma-ray emitter \citep{LHAASO:2025ysm}.

For sources that have exhibited super-Eddington activity, the column $\xi_{\rm Edd}$ reports the quantity $\xi_{\rm Edd}=L/L_{\rm Edd}$, where $L$ is the observed maximum X-ray (bolometric, when available) luminosity of the source and $L_{\rm Edd}$ is the corresponding Eddington luminosity.
%
For SS~433 and Cygnus X-3, as the nature and the mass of the compact objects are uncertain, also the estimated Eddington luminosity is uncertain.
For the former, we adopt the value $M_{\rm BH}=4.6 ~M_{\odot}$ as reported by \cite{2018Gaia}. For the latter, we need to make some considerations. According to \cite{Veledina:2023zho}, the minimum X-ray bolometric luminosity of Cygnus X-3 is $3\times 10^{38}\, \rm erg \, s^{-1} $. 
This luminosity can be up to a factor of 10 higher\footnote{This is true if we assume that the maximum value of the half-opening angle of the funnel allowed by X-ray polarization measurements is $\sim 15^{\circ}$  \citep{Veledina:2023zho}.}.
Assuming a black hole mass of $M_{\rm BH}=5~M_{\odot}$, the Eddington luminosity for a helium-dominated object like Cygnus X-3 is $L_{\rm Edd}\sim 1.25\times 10^{39}\, \rm erg \, s^{-1} $.
Comparing this with the bolometric luminosity, we estimate $\xi_{\rm Edd}\sim2.4$, as reported in Table \ref{table sources}.
It is important to note that a different assumption for the black hole mass will automatically translate into a different value of $\xi_{\rm Edd}$; however, this does not affect our results, since they depend on the product of $\xi_{\rm Edd}L_{\rm Edd}$ (see Sec.\ref{sec: Grammage}).
%
In the last column of Table \ref{table sources}, we report the gamma-ray energy flux at 100 TeV as given by \cite{LHAASOmicroquasars,LHAASO:cygnus}.
Regarding SS~433, the reported value corresponds to the flux at 100 TeV of the total emission of this source as derived from our best-fit power-law model obtained by fitting the data above $50$ TeV (see Appendix \ref{app: spectrum}).

\begin{table*}[ht]
\caption{The first column reports the source name, the second and the third columns the distances $d$ expressed in kpc and masses of the compact object $M_{\rm BH}$ expressed in solar masses \citep{BlackCAT}, respectively. The fourth column represents the extension of the gamma-ray emission $R_{\rm s}$ expressed in pc. For the sources that have shown super-Eddington activity, we also report in the fifth column the value $\xi_{\rm Edd}=L/L_{\rm Edd}$, with $L$ the observed maximum luminosity of the source and $L_{\rm Edd}$, the Eddington luminosity. The sixth column reports the gamma-ray flux at 100 TeV as measured by the LHAASO collaboration.}
\label{table sources}
\centering
\begin{tabular}{l|ccccc}
\hline
\hline
Source name & $d(kpc)$ & $M_{\rm BH}(M_\odot)$& $R_{\rm s}(pc)$ & $\xi_{\rm Edd}$ & $\phi_{\gamma,100 \, \rm TeV}(\rm  erg \, cm^{-2} s^{-1})$ \\
\hline
MAXI~J1820+070   & $2.96$\tablefootmark{a} & $8.4$\tablefootmark{b} & $14$ & $-$ & $2\times 10^{-14}$ \\
GRS~1915+105 & $9.4$ \tablefootmark{c} & $12$ \tablefootmark{c} & $45.9$ & $2-3$ \tablefootmark{d} & $1.1\times 10^{-13}$\\
V4641~Sgr & $6.2$ \tablefootmark{e} & $6.4$ \tablefootmark{e} & $35.7$ & $300$  \tablefootmark{f}& $2.6\times 10^{-12}$\\
SS~433  & $4.6$ \tablefootmark{g} & $4.2$\tablefootmark{h}  & $25.7$ & $6$ \tablefootmark{i}  & $1.6\times 10^{-13}$\\
Cygnus X-3 & $9$ \tablefootmark{c}  & $5$ & $13.5$ & $2.4$  & $1.0\times 10^{-13}$\\
\hline
Cygnus X-1  & $2.2$\tablefootmark{j}  & $21.2$\tablefootmark{j} & $<8.4$ & $-$  & $<1.0\times 10^{-14}$\\
XTE~J1859+226  & $12.5$\tablefootmark{k} & $8$ \tablefootmark{l} & $-$ & $-$ & $-$\\
GS~2000+251  & $2.7$\tablefootmark{m} & $5.5$\tablefootmark{n}  & $-$ & $-$ & $-$\\
GRO~J0422+32   & $2.49$\tablefootmark{o}  & $15$\tablefootmark{p}& $-$& $-$ & $-$ \\
V404~Cyg   & $2.39$\tablefootmark{q}  & $9$ \tablefootmark{r} & $-$ & $2$ \tablefootmark{s} & $-$\\
XTE~J1118+480 & $1.72$\tablefootmark{t}  & $6.9$\tablefootmark{u} & $-$ & $-$ & $-$ \\
V616   & $1.06$\tablefootmark{v} & $6.6$\tablefootmark{z}  & $-$ & $-$ & $-$\\
\hline
\end{tabular}
\tablefoot{ \tablefoottext{a}{\cite{Atri:2019kws}}, 
\tablefoottext{b}{\cite{Torres:2020ijh}}, 
\tablefoottext{c}{\cite{Reid:2023ksq}}, 
\tablefoottext{d}{\cite{Reid:2014ywa}}, 
\tablefoottext{e}{\cite{2014ApJ...784....2M}}, 
\tablefoottext{f}{\cite{Revnivtsev:2001hk}}, 
\tablefoottext{g}{\cite{2018Gaia}}, 
\tablefoottext{h}{\cite{Kubota:2010}}, 
\tablefoottext{i}{\cite{2009MNRAS.398.1668O}},
\tablefoottext{j}{\cite{2021Sci...371.1046M}},
\tablefoottext{k}{\cite{2011MNRAS.413L..15C}},
\tablefoottext{l}{\cite{Rizo:2022bma}},
\tablefoottext{m}{\cite{2004MNRAS.354..355J}},
\tablefoottext{n}{\cite{2004AJ....127..481I}},
\tablefoottext{o}{\cite{2003ApJ...599.1254G}},
\tablefoottext{p}{\cite{1997MNRAS.290..303B}},
\tablefoottext{q}{\cite{2009ApJ...706L.230M}},
\tablefoottext{r}{\cite{2010ApJ...716.1105K}},
\tablefoottext{s}{\cite{Motta:2017mol}},
\tablefoottext{t}{\cite{2006ApJ...642..438G}},
\tablefoottext{u}{\cite{2013AJ....145...21K}},
\tablefoottext{v}{\cite{2010ApJ...710.1127C}},
\tablefoottext{z}{\cite{2010ApJ...710.1127C}}.
}
\end{table*}

\section{Method}
\label{sec: Method}

\subsection{Transport scenario}
\label{sec: Transport}

We consider a two-zone transport model for CRs from microquasars.
Close to the source, the diffusion coefficient is reduced compared to the Galactic value, while further away particles propagate with the standard Galactic coefficient.
This scenario is motivated by recent observations of different classes of bright non thermal sources around which the diffusion coefficient appears to be significantly suppressed compared to the Galactic one, including pulsar wind nebulae \cite[see, e.g.,][]{HAWC:2017kbo}, supernova remnants \citep{Loru+2021, MAGIC_gamma-Cyg:2023}, and  star clusters \citep{LHAASO:cygnus}. 
The suppression of the diffusion coefficient may be due to either a local enhancement of magnetic turbulence or to the self-generation of turbulence driven by escaping CRs \citep{Evoli-Linden_Morlino:2018, Schroer+2022}. In the case of microquasars, the turbulence enhancement may be also connected to the interaction between the jet and the surrounding medium, which results in a highly turbulent and pressurized cocoon.

We calculate the CR number density $n(E,r,t)$ by numerically solving the transport equation in spherical symmetry:
\begin{equation}
\label{eq: transport}
\frac{\partial n(E,r,t)}{\partial t}=
\nabla \cdot \left[D(E,r) \, \nabla n(E,r,t)\right] + Q(E) \, \delta^{3}(\vec{r})
\end{equation}
where $Q(E)$ is the injection rate and $D(E,r)$ is the diffusion coefficient.  
We assume continuous injection over a duration of $0.1-1 \, \rm Myr$ with the injection rate defined as:
\begin{equation}
Q(E)=Q_{0} \, g(E)
\end{equation}
where $g(E)$ is a power-law with an exponential cutoff, characterized by a spectral index $\alpha=2$ and cutoff energy $E_{\rm cut}=1$ PeV (see also Appendix~\ref{app: spectrum} for a comparison between the gamma-ray flux predicted from this assumed proton spectrum and the observed gamma-ray data). 
The normalization $Q_0$ is related to the source luminosity as
\begin{equation}
 Q_0= \eta_{\rm CR} \xi_{\rm Edd}\frac{L_{\rm Edd}} {\int_{1 \, \rm GeV}^{10 \, \rm PeV} g(E) E dE}= \eta_{\rm CR} \xi_{\rm Edd}Q_0'
\end{equation}
where $\eta_{\rm  CR}$ is the CR acceleration efficiency, which is fixed to $0.1$, $L_{\rm Edd}$ is the Eddington luminosity, and $\xi_{\rm Edd}>1$ allows for super-Eddington bolometric luminosities.

The diffusion coefficient factorizes as $D(E,r)=f(r) K(E)$, where the energy dependence is assumed to be the same inside and outside the suppressed region, i.e., $K(E)=(E/E_0)^{\delta}$, with $E_0=100$ TeV and $\delta=0.339$.
We obtain $\delta$ by approximating the high-energy parameterization ($E>1$ TeV) of the diffusion coefficient proposed by \cite{Evoli:2020szd} with a simple power-law, neglecting the low-energy break since our analysis is restricted to energies above it.
The spatial dependence is:
\begin{equation}
f(r)=\begin{cases}
D_0, & r \le r_0 \\
D_1, & r > r_0
\end{cases}
\end{equation}
where $D_0$ and $D_1$ are the diffusion coefficients near and far from the source, respectively. 
We adopt $D_1=3.5\times 10^{30} \, \rm cm^{2} \, s^{-1}$, obtained from the same high-energy approximation to the diffusion coefficient of \cite{Evoli:2020szd}, and $D_0=D_1/100$ \citep{LHAASOmicroquasars}.
The size of the suppressed diffusion region, $r_0$, in our reference case, is set to $\sim 50$ pc, which is comparable to the largest gamma-ray extension observed from microquasars detected by LHAASO,  this choice is discussed in more details in Section \ref{sec: Grammage}.

\begin{figure}[ht]
\includegraphics[width=0.5\textwidth]{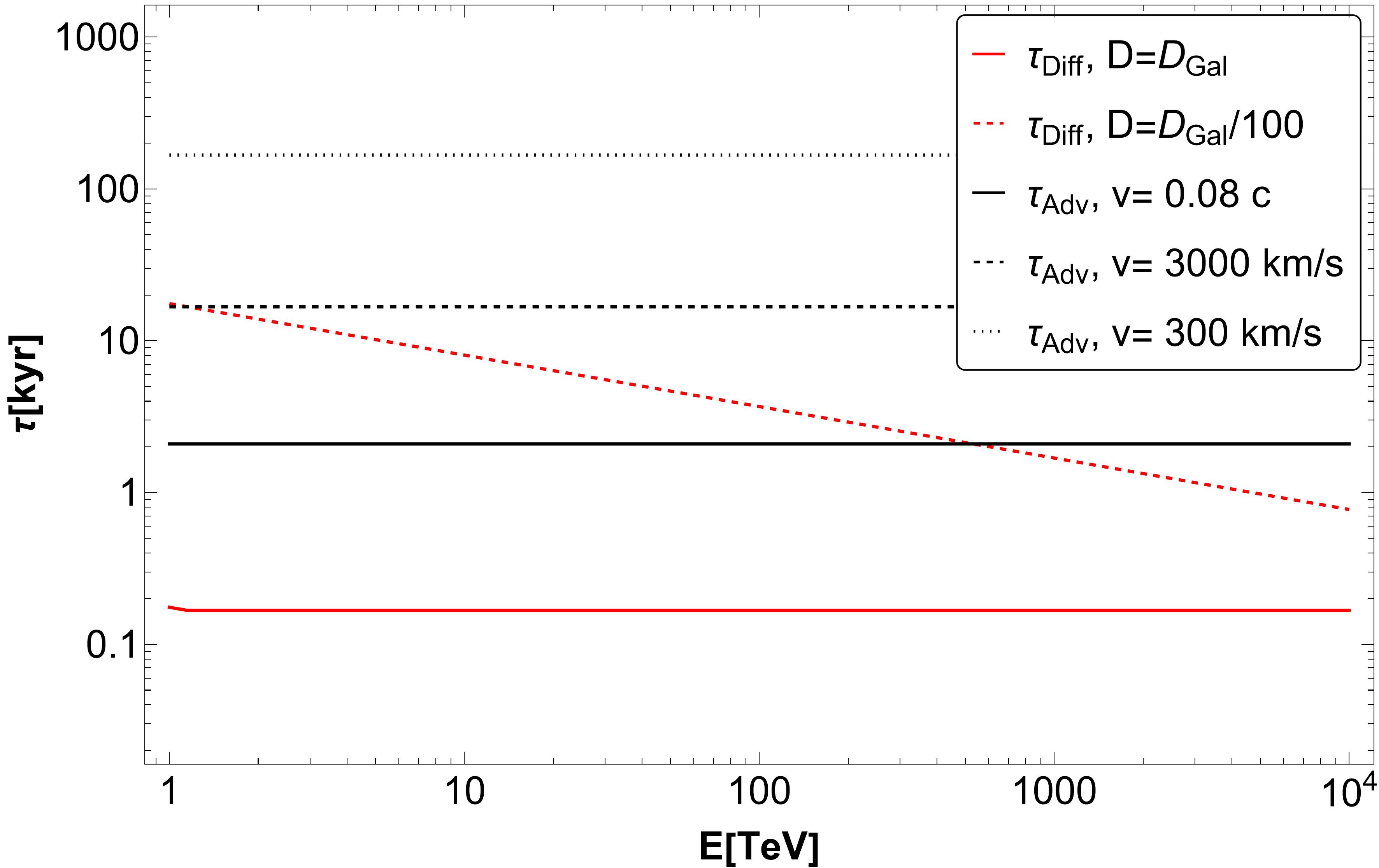}
\caption{ Comparison of diffusion and advection timescales over a region of $50$ pc. The red solid (dashed) line shows the diffusion timescale for the Galactic (suppressed) diffusion coefficient.
The black solid, dashed, and dotted lines represent the advection timescales for an advection velocity of $v=0.08 \, c$ (comparable to the jet velocity) and for wind velocity $v=300 \, \rm km \, s^{-1}$ and $v=3000 \, \rm km \, s^{-1}$, respectively.}
\label{fig: timescales}
\end{figure}

Microquasars may host equatorial winds with velocities in the range $\sim300-3000 \rm \, km \, s^{-1}$  \cite[see, e.g.,][]{Ponti:2012}.
However, we neglect the advection term in the transport equation  within the region of 50 pc, since, for both the standard and the suppressed diffusion assumed here, the diffusion timescales are always shorter than the advection timescales at energies above 100 TeV (see Figure \ref{fig: timescales}).
Notice that when the Galactic diffusion coefficient, $D(E)=D_1 K(E)$, is adopted for a region of size $R=50$ pc, the diffusion approximation implies superluminal propagation at all energies considered. Hence, the diffusion time-scale is calculated as $\tau_{\rm Diff}=\max[R^2/(6 \, D(E)), R/c]$) in order to avoid nonphysical super-luminal propagation. 


%
Advection could instead dominate over diffusion within the jet, where the bulk velocity can reach a non-negligible fraction of the speed of light. For example, in SS~433, $v \simeq 0.08 \, c$, consistent with the jet velocity measured by H.E.S.S. \citep{HESS:2024rlh}.
However, we neglect the jet in the present analysis because it is under-dense and therefore we do not expect a significant amount of hadronic emission from it.

%
Finally, the CR spectrum at Earth is calculated as:
\begin{equation}
    \phi_{\rm CR}= \frac{c}{4 \pi}n(E,r,t_{\rm age})
\end{equation}
where $r$ is the distance between the source and Earth, and $t_{\rm age}$ represents the age of the source, which in our model is equal to the injection duration, assumed to be in the range of $0.1-1 \, \rm Myr$.
The two-zone diffusion model considered here is better treated in spherical symmetry. However, this complicates the implementation of a free escape boundary at the edge of the Galactic magnetized halo, $H$, where CRs are usually assumed to escape \cite[see, e.g.,][]{Blasi_2012}. 
Proper treatment of this boundary is particularly relevant for sources injecting particles over long timescales whose propagation length becomes comparable to the size of the Galactic halo, which we assume to be $H=5 \rm \, kpc$ \citep{Evoli:2020szd}.

To account for the effect of the free escape boundary, we estimate the energy-dependent suppression factor using the analytical solution of the transport equation in Cartesian coordinates for a one-zone model with a diffusion coefficient equal to the Galactic value, $D(E)=K(E)D_1$ (see Appendix \ref{app: halo}).  
The derived energy dependent factor is then applied as a correction to the numerical solution obtained in the spherical two-zone model.
The suppression of the flux due to the finite halo size in this one-zone model provides a good approximation in our case, since the size of the region around the source where transport is suppressed with respect to the Galactic average, $r_0$, is much smaller than $H$. 
%

\subsection{Target density}
\label{sec: Grammage}

In this section we derive the target density required to reproduce the observed gamma-ray emission and calculate the grammage accumulated by CRs in the vicinity of the source.

The gamma-ray flux at Earth produced by $p-p$ interactions is given by:
\begin{equation}
\phi_{\gamma}(E_{\gamma})=\frac{n_{\rm T} \, c}{4 \pi d^2}\int_{0}^{R_{\rm s}}d^3r \int_{E_{\gamma}}^{\infty}dE' \, n(E',r, t_{\rm age}) \frac{d \sigma(E',E_{\gamma})}{dE_{ \gamma}} 
\label{eq: gamma1}
\end{equation}
where $d$ is the source distance, $n_{\rm T}$ is the target number density, and $\frac{d \sigma(E,E_{\gamma})}{dE_{ \gamma}}$ is the differential cross section, for which we adopt the parametrization provided by \cite{Kelner:2006tc}.
The spatial integration is performed over a spherical region of radius $R_{\rm s}$ centered on the source, corresponding to the extension of the gamma-ray emitting region.

We relate the hadronic gamma-ray flux to the observed flux as $\phi_\gamma=\eta \, \phi_{\gamma,\rm obs}$, where $\phi_{\gamma,\rm obs}$ is the gamma-ray flux measured by LHAASO and $\eta\le1$ represents the fraction of the observed emission attributed to hadronic interactions. 
As a first guess, we assume $\eta=1$.

By inverting Eq. \eqref{eq: gamma1}, we derive the target density material required to reproduce the observed gamma-ray flux $\phi_{\gamma,\rm obs}$ as a function of the size of the emission region $R_{\rm s}$, assuming the CR acceleration efficiency $\eta_{\rm CR}=0.1$.
The inferred density is then compared with the measured gas density in the source environment (see Sec. \ref{sec: measured gas}).

For the numerical estimates presented in this work, we adopt $R_{\rm s}= 50$ pc corresponding approximately to the largest gamma-ray extension reported by the LHAASO collaboration (see column $4$ of Table \ref{table sources}).
If the density required to explain the gamma-ray emission exceeds the measured value, the Eddington ratio $\xi_{\rm Edd}$ is rescaled accordingly.
The inferred values of $\xi_{\rm Edd}$ are subsequently compared with the measured values reported in Table \ref{table sources}.
When the required $\xi_{\rm Edd}$ exceeds the measured one, we derive an upper limit on $\eta$, defined as the maximum fraction of the observed gamma-ray flux that can be attributed to hadronic interactions.

We additionally derive the grammage accumulated by CRs in the vicinity of the source.
Assuming continuous particle injection over $t_{\rm age}$, the grammage can be inferred directly from the measured gamma-ray flux and the injected CR spectrum. 
We follow the approach of \cite{Blasi:2025yjl}.
For continuous injection over timescales of $0.1-1$ Myr and CRs with energies above 100 TeV within $r<r_0$, we can safely consider the stationary solution of the transport equation.
For example, assuming $r_0=50$ pc, CRs with energies of 100 TeV spend on average $\tau_{\rm Diff}=r_0^2/(6 D_0)\sim 4 \rm \, kyr$ inside the inner region with suppressed diffusion before entering the outer zone characterized by the Galactic diffusion coefficient.
This timescale is almost two orders of magnitude shorter than the minimum injection timescale considered in this work, namely $100$ kyr.
All the microquasars listed in Table \ref{table sources} that are detected by LHAASO are observed above $10$ TeV, with the exception of SS~433, which is detected above $\sim 1$~TeV. 
Gamma rays with energies of $10$ TeV are predominantly produced by CRs with characteristic energies around $100$ TeV, further justifying the stationary approximation.
In the case of SS~433, the approximation remains valid because the residence time of CRs of 10 TeV inside the suppressed-diffusion region is approximately $\sim 8$ kyr, still significantly shorter than the assumed injection timescale.

In the diffusion-dominated scenario considered in this work, assuming suppressed diffusion around the source and neglecting the boundary at $r=r_0$, the stationary solution of the transport equation is given by
\begin{equation}
n_{\rm st}(E,r)=\frac{Q(E)}{4 \pi r D_0 K(E)}\ .
\label{eq: stationary}
\end{equation}
This solution differs from the exact solution of Eq.~\eqref{eq: transport}, which properly accounts for the transition from suppressed diffusion to Galactic diffusion at $r \le r_0$ (see Appendix \ref{app: stationary vs solution}).
In particular, the exact solution systematically falls below the stationary solution near the boundary of the suppressed-diffusion region.

%
In order to calculate the grammage, we therefore approximate the numerical solution inside the suppressed-diffusion region with an analytical expression:
\begin{equation}
 n_{\rm approx}(E,r)=n_{\rm st}(E,r) \, \lambda \left(\frac{r}{r_0} \right) \rm 
\label{eq: approx solution}
\end{equation}
where 
\begin{equation}
 \lambda(x)= \rm Erfc \left[ a \, x^{b+c \,x} \right]
\end{equation}
with the parameters $a=1.28$, $b=1.09$ and $c=0.72$ fitted on the numerical solution.

This approximation becomes less accurate near $r_0$, but remains sufficiently accurate for our purposes and has only a minor impact on the predicted gamma-ray flux. 
%
In particular, the volume integral of the approximate solution agrees with that of the full numerical solution within $3\%$ for all energies above $10$ TeV when $R_{\rm s}=r_0$.


%
The grammage is defined as:
\begin{equation}
X_{\rm Diff}(E)=n_{\rm T} m_{\rm p}c \tau_{\rm Diff}(E)=\frac{X_{0}}{K(E)}
\label{eq: grammage}
\end{equation}
where the diffusion timescale is given by $\tau_{\rm Diff}(E)=R_{\rm s}^2/(6\,D_0\, K(E))$.
Here $X_0$ denotes the grammage at the reference energy $E_0= 100$ TeV.

Substituting the approximate solution, Eq.~\eqref{eq: approx solution}, into Eq.~\eqref{eq: gamma1}, the gamma-ray flux can be expressed as:
\begin{equation}
\phi_{\gamma, \rm Diff}(E_{\gamma})=\frac{1}{4 \pi d^2} \frac{X_{0}}{m_{\rm p} } Q_0 \, F_{\rm Diff}(E_{\gamma}) \, C(r_0, \,R_{\rm s}),
\label{eq: gamma2}
\end{equation}
where
\begin{equation}
C(r_0,R_{\rm s})= 2 \left(\frac{r_0}{R_{\rm s}}\right)^2\int_{0}^{\frac{R_{\rm s}}{r_0}} dx \, x \, \lambda(x)
\label{eq: factor}
\end{equation}
with $R_{\rm s}\le r_0$, and
\begin{equation}
F_{\rm Diff}(E_{\gamma}) =3\int_{E_{\gamma}}^{\infty}dE' \, \frac{g(E')}{K(E')} \frac{d \sigma(E',E_{\gamma})}{dE_{ \gamma}} \,.
\end{equation}
The grammage normalization required to reproduce the observed gamma-ray flux is then given by
\begin{equation}
X_{0}=  \frac{4 \pi d^2m_{\rm p}}{\eta_{\rm CR} \xi_{\rm edd} Q_0' F_{\rm Diff}(E_\gamma)C(r_0, R_{\rm s})} \, \phi_{\gamma}(E_{\gamma})
\label{eq: grammage 2}
\end{equation}
where $\phi_{\gamma}=\eta \,\phi_{\gamma,\rm obs}$.
 In this work we adopt $r_0=R_{\rm s}$, which yields $C\sim 0.35$.


\subsection{Measured gas density}
\label{sec: measured gas}
To estimate the gas density around the microquasars, we make use of data cubes of CO line emission, which trace molecular H$_2$ gas \citep{2001ApJgasCO}, and of HI brightness temperature \citep{2016A&AgasHI}. 
These cubes are constructed in bins of Galactic longitude, latitude, and radial velocity, where the latter corresponds to the projection of the Galactic rotation velocity along the line of sight. Since the rotation velocity depends on the Galactocentric distance $R$, the measured radial velocity can be related to the gas distance, although with some uncertainty. In this work, we adopt the Galactic rotation curve $V(R)$ parametrized by \cite{2019ApJrotationCurve}.

Given the distance $d$ of each microquasar, we determine its Galactocentric radius $R_\ast$ and the corresponding rotation velocity $V_\ast = V(R_\ast)$. We then select gas within a velocity interval $\Delta V$ around $V_\ast$, corresponding to a spatial region surrounding the source. The interval $\Delta V$ is defined by assuming characteristic source radii $R_{\rm s} = 50$ and $100$ pc. Specifically, we compute the velocities $V_1$ and $V_2$ corresponding to distances $d - R_{\rm s}$ and $d + R_{\rm s}$, respectively.

The gas density is then estimated by integrating the emission within a cylindrical volume centered on the source, with radius $R_{\rm s}$ and height $2R_{\rm s}$ along the line of sight, corresponding to the velocity interval between $V_1$ and $V_2$. The integrated intensities are converted into gas column densities using the standard conversion factors $X_{\rm CO} = 4\times10^{20}\,{\rm cm^{-2}\,(K\,km\,s^{-1})^{-1}}$ for molecular gas and $X_{\rm HI} = 1\times10^{18}\,{\rm cm^{-2}\,(K\,km\,s^{-1})^{-1}}$ for atomic hydrogen.
We take into account the heavy elements by assuming the same composition of the solar system and multiplying the cubes by a factor 1.42 \citep{Ferriere:2001rg}.

For V4641~Sgr and MAXI~J1820+070, the gas maps considered in this work do not completely cover the source region; therefore, the resulting target gas density should be regarded as a lower limit.
An upper limit on the gas number density within the gamma-ray emission region of V4641~Sgr is reported by \cite{HESS:2025uyo}, who estimate $n_{\rm T}\leq 0.2 \, \rm cm^{-3}$.


\begin{figure}[ht]
\includegraphics[width=0.45\textwidth]{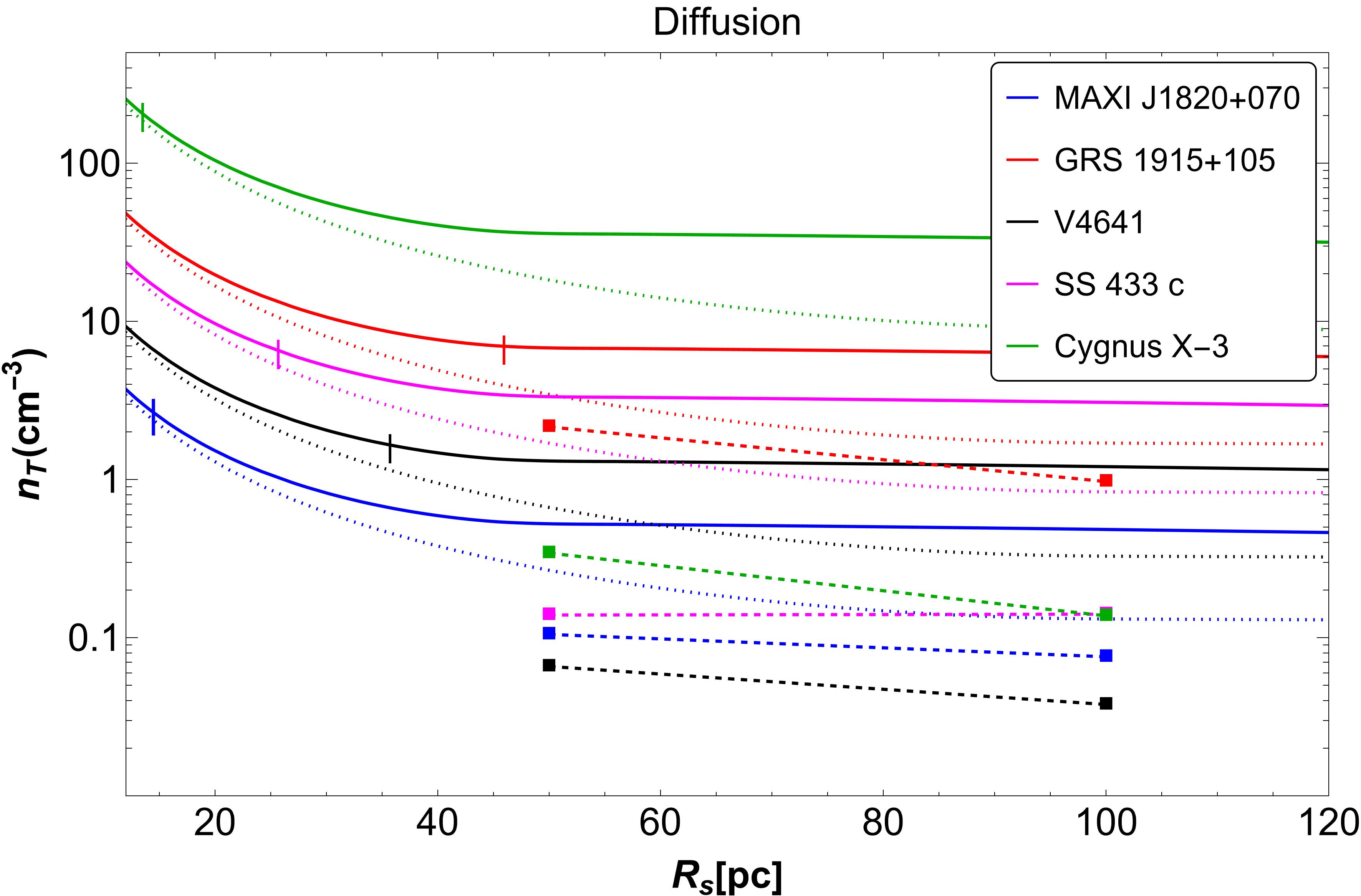}
\caption{Estimated target density required to reproduce the gamma-ray emission at 100 TeV as a function of $R_{\rm s}$, the radius of the interaction region, for the microquasars detected by LHAASO. The estimates are shown as solid (dotted) lines for $r_0=50$ pc ($r_0=100$ pc). They are obtained assuming $\xi_{\rm Edd}$ as in Table \ref{table sources} (when $\xi_{\rm Edd}$ is not reported we assume the value 1), $\eta_{\rm CR}=0.1$ and considering continuous injection. The measurements of the gas number density within $50 \, \rm pc, --100 \, \rm pc$ are shown as filled squares connected by dashed lines to guide the eye. We also report the $1 \sigma$ extension of the gamma-ray emission region $R_{\rm s}$ as a bar.}
\label{fig: number density detected}
\end{figure}

\begin{figure*}[ht]
\centering
\includegraphics[width=0.45\textwidth]{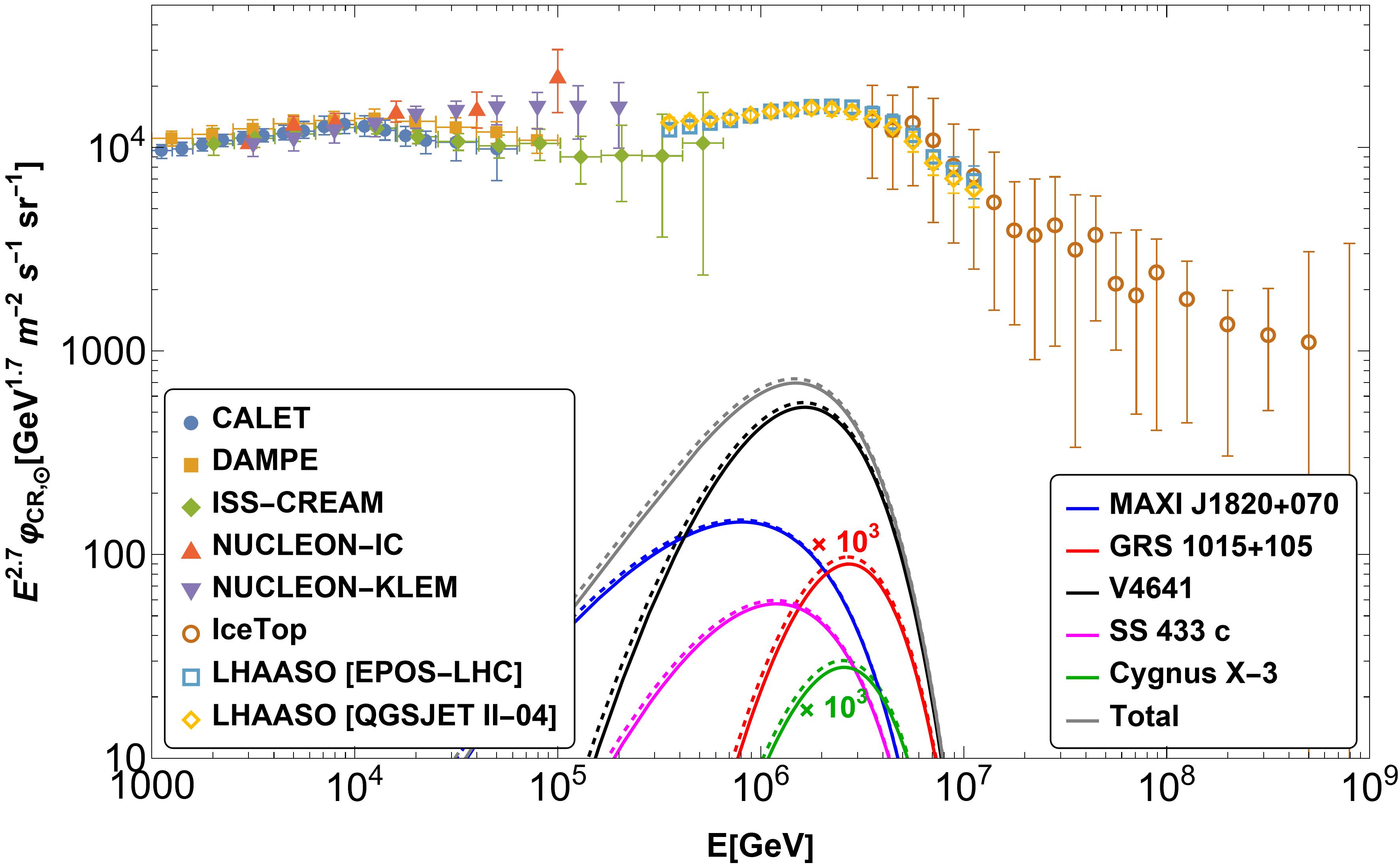}
\includegraphics[width=0.45\textwidth]{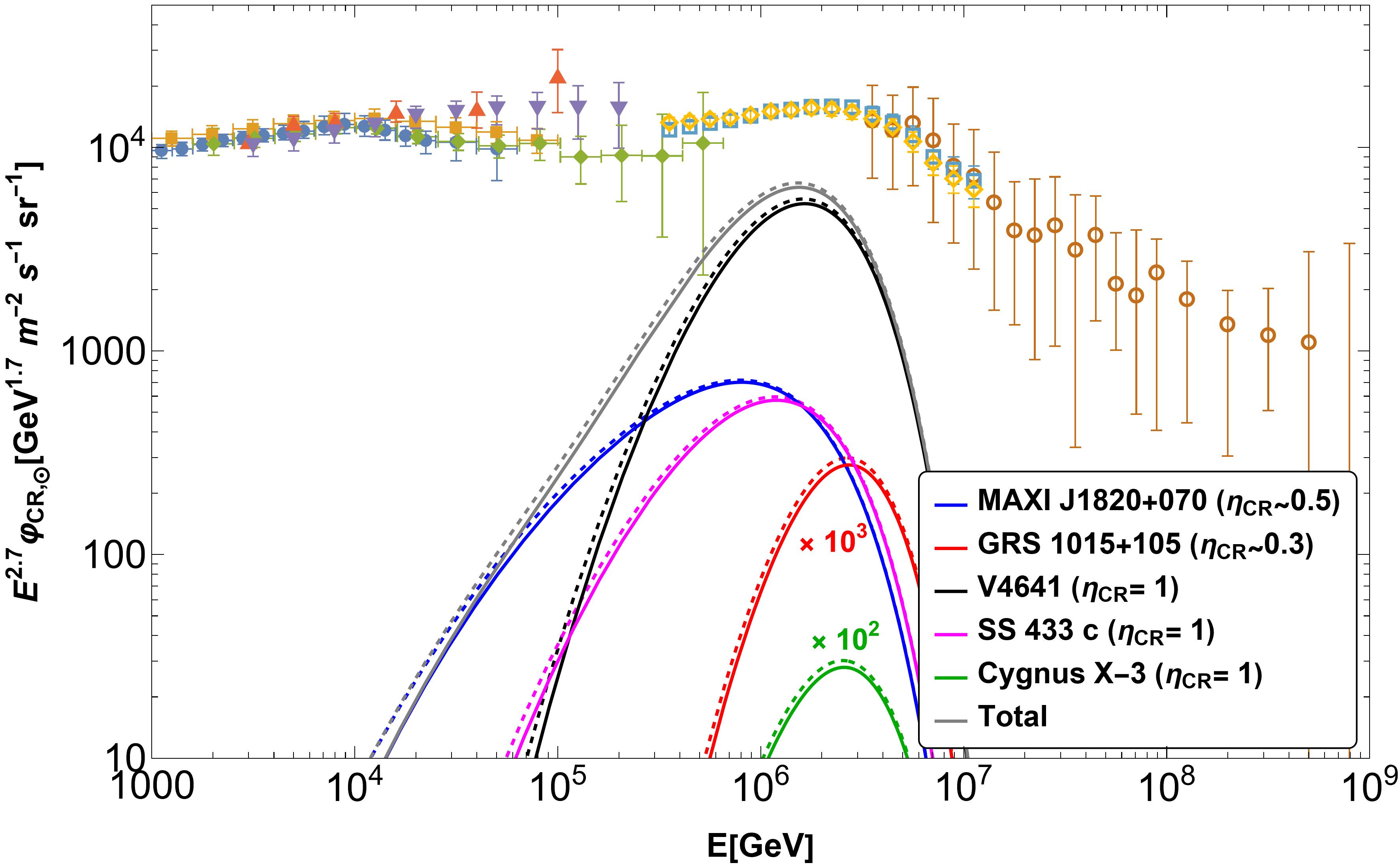}
\caption{Contribution to the CR proton spectrum for the microquasars detected by LHAASO in the diffusion-dominated two-zone model assuming continuous injection over 100 kyr (solid lines), compared with proton data from LHAASO \citep{LHAASOprotons}, IceTop \citep{IceCube:2019hmk}, CALET \citep{CALET:2022vro}, DAMPE \citep{DAMPE:2019gys}, ISS-CREAM \citep{2022ApJISS-CREAM}, NUCLEON-IC, and NUCLEON-KLEM \citep{2019AdSpR_NUCLEON}.
The contribution expected from a one-zone model with the Galactic diffusion is shown as dashed lines. 
%
The left and right panels show the results for standard case with $\eta_{\rm CR}=0.1$, and for the maximum-efficiency case, in which $\eta_{\rm CR}$ is adjusted to reproduce the observed gamma-ray flux at 100 TeV when possible.
The fluxes of MAXI~J1820+070 and Cygnus X-3 are multiplied by a factor $10^3$ in the left panel, and by a factor $10^3$ and $10^2$ respectively in the right panel. This rescaling reduces the dynamic range of the y-axis so as to improve readability.
}
\label{fig: CR detected}
\end{figure*}

\begin{figure}
\includegraphics[width=0.45\textwidth]{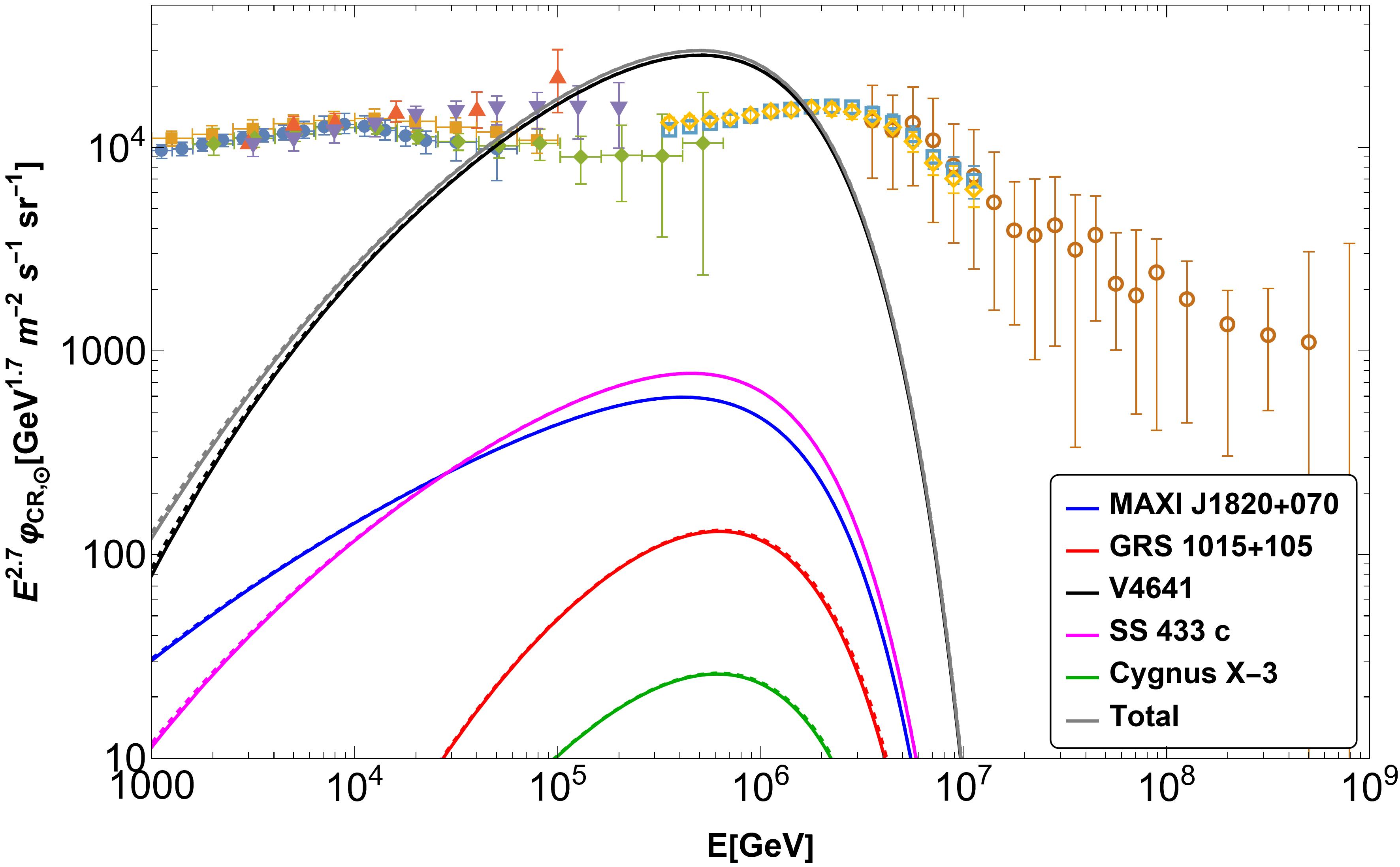}
\caption{Same as Figure \ref{fig: CR detected} for our standard case $\eta_{\rm CR}=0.1$ and continuous injection over 1 Myr.}
\label{fig: CR detected 1 Myr}
\end{figure}

\begin{table*}[ht]
\caption{For each microquasar detected by LHAASO, we report the grammage $X_{\rm Diff}$ accumulated at proton energy of $1$ PeV ($E_{\gamma}\sim100$ TeV), the Eddington ratio $\xi_{\rm Edd}$ required to reproduce the observed gamma-ray flux at 100 TeV assuming an interaction region of radius $R_{\rm s}=50\,\rm pc$ and $r_0=R_{\rm s}$, the maximum fraction $\eta$ of the gamma-ray flux that can be attributed to hadronic interactions, and the contribution $\omega$ to the CR spectrum at 1~PeV for sources emitting continuously over 100 kyr and 1 Myr, expressed as a fraction of the observed CR flux at that energy at Earth.
All quantities are computed assuming $\eta_{\rm CR}=0.1$.}
\label{tab:diffusion}
\centering
\begin{tabular}{lccccc}
\hline\hline
Source &
$X_{\rm Diff}(1\,\mathrm{PeV})$ [$\mathrm{g\,cm^{-2}}$] &
$\xi_{\rm Edd}(R_{\rm s}=50\,\mathrm{pc})$ &
$\eta$ &
$\omega_{\rm Diff}$ (0.1 Myr) &
$\omega_{\rm Diff}$ (1 Myr) \\
\hline
MAXI~J1820+070 & $1.3\times10^{-3}\,\xi_{\rm Edd,1}^{-1}$ & $4.8$ & $20\%$ & $1\%$ & $3.2\%$ \\
GRS~1915+105   & $1.7\times10^{-2}\,\xi_{\rm Edd,3}^{-1}$ & $9.2$ & $32\%$ & $\ll1\%$  & $0.8\%$ \\
V4641~Sgr          & $3.3\times10^{-3}\,\xi_{\rm Edd,300}^{-1}$ & $5.8\times10^{3}$ & $5\%$ & $2.9\%$ & $163\%$ \\
SS~433         & $8.4\times10^{-3}\,\xi_{\rm Edd,6}^{-1}$ & $140$ & $4\%$ & $0.4\%$ & $4.3\%$ \\
Cygnus X-3 & $4.5\times10^{-2}\xi_{\rm Edd,2.4}^{-1}$ & $122$ & $2\%$ & $\ll1\%$ & $0.1\%$ \\
\hline
$\sum$       & $0.07<0.4$ & - & $-$ & $4.3\%$ & $171\%$ \\
\hline
\end{tabular}
\end{table*}

\section{Results}
\label{sec: Results}

\subsection{Detected microquasars}
In the second column of Table \ref{tab:diffusion}, we report the grammage required to reproduce the observed gamma-ray flux at 100 TeV for the five microquasars detected by LHAASO at this energy, assuming that the fraction of the observed gamma rays produced by hadronic interactions is $\eta=1$.
The grammage is computed using Eq.~\eqref{eq: grammage 2}, assuming $r_0=R_{\rm s}=50$ pc,  $\eta_{\rm CR}=0.1$ and adopting measured values of $\xi_{\rm Edd}$ when available (Table~\ref{table sources}); otherwise, we assume $\xi_{\rm Edd}=1$.
%
The grammage required to account for the detected UHE gamma-ray emission as hadronic never exceeds the best-fit value for the so-called source grammage, $X_s=0.4 \, \rm g \, cm^{-2}$, as constrained by fits to the $B/C$ (Boron over Carbon flux) observations \citep{Evoli:2019wwu}. 
This provides a consistency check for scenarios invoking the UHE detected microquasars as the primary contributors to PeV CRs: the hadronic interpretation of their UHE gamma-ray emission implies a locally accumulated grammage that does not violate any constraint.
%

%
Assuming the same values of $\xi_{Edd}$, $\eta_{\rm CR}$, and $\eta$ as in the grammage calculation, we estimate the target number density required to produce the gamma rays for $r_0=50$ pc and $r_0=100$ pc.
In Figure \ref{fig: number density detected}, we show the estimated target number density as a function of the radius of the gamma-ray emission region together with measurements of CO and HI in the same regions (see Sec.\ref{sec: measured gas}).
The required target density (solid lines for $r_0=50$ pc and dotted lines for $r_0=100$ pc) decreases with increasing $R_s$ and flattens for $R_{\rm s}\ge r_0$. This occurs because the CR density drops by a factor of $\sim 100$ at the boundary between the two diffusion zones. As a consequence, the gamma-ray emission is dominated by interactions occurring within $R_{\rm s}\le r_0$ where the CR density is much higher.

From Figure \ref{fig: number density detected}, it is clear that the average values of the measured gas density within regions of size 50 and 100 pc around the sources are not sufficient to account for the full observed gamma-ray flux at 100 TeV through $p-p$ interactions for any of the detected microquasars.
From the comparison between the theoretical prediction and the gas measurements at $R_{\rm s}=50$ pc (corresponding to the largest measured extent of the gamma-ray emission), we estimate the values of $\xi_{\rm Edd}$ required to explain the total observed gamma-ray emission at 100 TeV. 
These values are reported in the third column of Table \ref{tab:diffusion} for the case of suppressed diffusion up to $r_0=50$ pc. 
We find that the derived $\xi_{\rm Edd}$ values exceed the measured ones reported in Table \ref{table sources} for all systems, clearly in tension with an interpretation of the gamma-ray emission as purely due to hadronic $p-p$ interactions.

%

%
This tension can be alleviated in several ways. 
One  possibility is to assume a larger acceleration efficiency $\eta_{\rm CR}$. In the case of MAXI~J1820+070 and GRS~1915+105, accounting for the observed UHE emission with the measured target density would require an efficiency $\eta_{\rm CR}=0.49$ and $0.31$, respectively. In all other cases $\eta_{\rm CR} >1$ would be required.

A complementary approach is to further reduce the diffusion coefficient $D_{0}$. To be fully compatible with the observed gamma-ray fluxes, the Galactic diffusion coefficient would need to be suppressed by factors of $\sim(486,\, 306 ,\, 1933, \, 2332, \, 5090)$ for MAXI~J1820+070, GRS~1915+105, V4641~Sgr, SS~433 and Cygnus X-3, respectively.
However, we stress that the diffusion coefficient cannot fall below the Bohm limit, $D_{\rm Bohm}(E)=E c/(3qB)$. By assuming $B=20\, \mu G$ for all microquasars \citep[consistent with the value derived for SS~433 by][]{HESS:2024rlh}, we can express the required value of the diffusion coefficient at 1 PeV, $D_0 K(1~ \rm PeV)$, as a multiple of $D_{\rm Bohm}(1\, \rm PeV)=5\times10^{27} \, \rm cm^{2} \, s^{-1}$.
We get $D_0 K(E)/D_{\rm Bohm}(E)|_{E=1 \rm PeV}  \sim(3, \, 5, \, 0.8, \, 0.6, \, 0.3)$ for MAXI~J1820+070, GRS~1915+105, V4641~Sgr, SS~433, and Cygnus X-3, respectively. Those values suggest that, under the assumption of a highly suppressed diffusion coefficient, only MAXI~J1820+070 and GRS~1915+105 are compatible with a purely hadronic origin of the observed emission.
%
It should be noted, however, that the magnetic field in microquasar jets, which affects the exact value of $D_{\rm Bohm}$, is very uncertain.

Finally, it is possible that the gamma-ray emission at 100 TeV is not purely hadronic.
Under the assumption $\eta_{\rm CR}=0.1$, we can therefore derive the maximum fraction of the gamma-ray flux that can be attributed to hadronic interactions, $\eta$ (see Table \ref{tab:diffusion}).
In general, the parameter $\eta$ is degenerate with the assumed CR efficiency $\eta_{\rm CR}$.
The fraction $\eta$ doubles at 100 TeV if the cutoff energy of the injected proton spectrum is increased from 1 PeV to $10$ PeV (see Appendix \ref{app: spectrum}).

The fraction $\eta$ could become slightly larger if we consider a larger size for the suppressed diffusion region, e.g., $r_0=100$ pc (see the dotted lines in Figure \ref{fig: number density detected}). In this case, if we still assume $R_{\rm s}= 50$ pc, $\eta$ is approximately a factor of $2$ larger than the value reported in Table \ref{tab:diffusion}.
In conclusion, while MAXI~J1820+070 and GRS 1915+105 can be reconciled with a predominantly hadronic interpretation,  SS~433, V4641~Sgr, and Cygnus X-3 are challenging to explain with purely $p$–$p$ interactions.

In the case of increased CR efficiency or a reduced hadronic fraction of the gamma-ray emission, the accumulated grammage would decrease with respect to the values reported in the tables,  while it would remain unchanged in the case of a further reduction of the diffusion coefficient (see Eq. \eqref{eq: grammage} and Eq. \eqref{eq: grammage 2}).

At this point we can proceed by estimating the maximum contribution of these sources to the CR “knee” for different assumptions on the source age. 
%
In Fig.~\ref{fig: CR detected}, we show the predicted CR proton spectra at Earth assuming continuous injection over $100$ kyr for each source (solid lines), compared with CR data.
The left panel corresponds to our benchmark case having CR acceleration efficiency $\eta_{\rm CR}=0.1$, while the right panel shows the maximum possible contribution to the CR proton spectrum.
In the latter case, we adopt the largest values of $\eta_{\rm CR}$ consistent with the observed gamma-ray emission at 100 TeV, namely $\eta_{\rm CR}=0.48$ for MAXI J1820+070, and $0.3$ for GRS 1915+105. 
For V4641~Sgr, SS~433, and Cygnus X-3, the efficiency required to explain the observed gamma-ray emission exceeds unity; we therefore adopt $\eta_{\rm CR}=1$. Even in this case, the contribution of Cygnus X-3 to the CR spectrum remains negligible due to its large distance from Earth.
Figure~\ref{fig: CR detected 1 Myr} shows the same calculation as done for the benchmark case but for a source duration of 1 Myr.

To evaluate the impact of the suppressed diffusion, we also show, with dashed lines, the results for a one-zone model, where the diffusion is everywhere identical to the Galactic one.
The effect of a reduced diffusion coefficient around sources is mainly to delay the escape of the lower-energy particles with respect to the one-zone model but does not affect the final flux for the energy range of our interest, even for a source age of 100 kyr. 
%
In general, in the case of the two zone model, for injection times over a period $t_{\rm age}\ll 100$ kyr the contribution of microquasars to the CR spectrum is limited by the suppressed diffusion region around the source, since CRs are likely still confined inside. On the other hand, for $t_{\rm age}\gg 100$ kyr, as in the case treated in this work of $t_{\rm age}=1$ Myr, the contribution to the CR spectrum is limited by the halo size, in particular for sources with distance $d>H$.

The contribution of LHAASO-detected microquasars to the CR proton spectrum strongly depends on the injection history.
If these sources continuously emit PeV CRs over $\sim100$ kyr, their contribution remains negligible (below $\sim5\%$ at $\sim 1$~PeV) in the benchmark case with $\eta_{CR}=0.1$, and can reach up to $37\%$ in the maximal-efficiency case.
If instead the emission persists for $\sim1$ Myr, even with $\eta_{\rm CR}=0.1$, the predicted flux from V4641~Sgr eventually overshoots the observed CR spectrum.
This suggests that V4641~Sgr cannot be active as a PeVatron over such a long timescale.

\subsection{Undetected microquasars}

\begin{figure}[ht]
\includegraphics[width=0.45\textwidth]{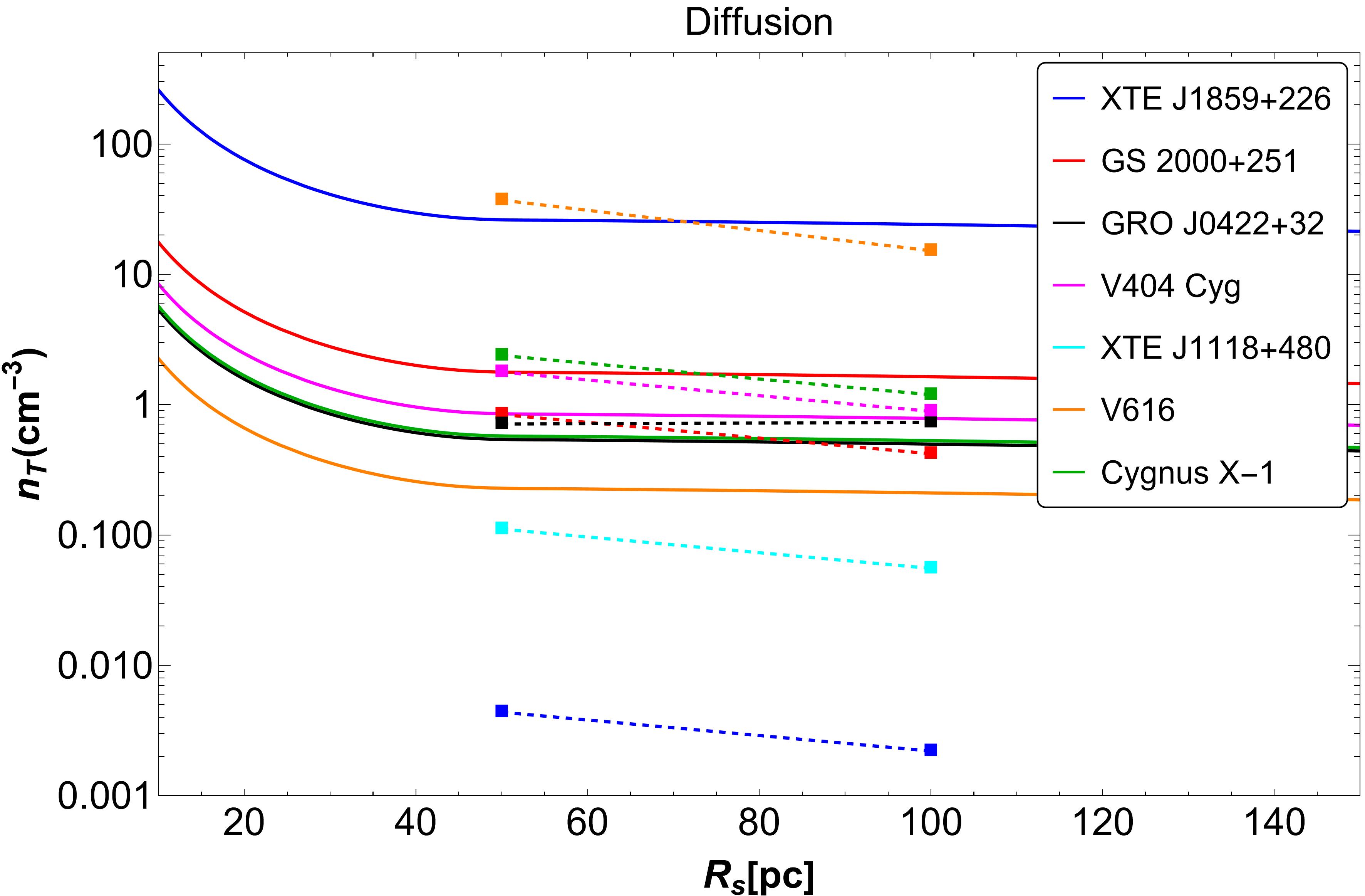}
\caption{Same as Fig.\ref{fig: number density detected} but for the microquasars undetected by LHAASO. The predictions for GRO J0422+32, XTE J1118+480 and Cygnus X-1 overlap.}
\label{fig: number density non-detected}
\end{figure}

A similar approach is applied to microquasars that are not detected by LHAASO.
In Figure \ref{fig: number density non-detected}, we show the estimated upper limits on the target density as a function of the gamma-ray emission radius, again compared with the total gas density as estimated from CO and HI measurements assuming $r_0=50$ pc.

The upper limits on the target density are derived assuming $\eta_{\rm CR}=0.1$ , $\xi_{\rm Edd}=1$ and requiring that the gamma-ray flux at 100 TeV satisfies $\phi_{\gamma}(100\, \rm TeV)\le\phi_{\rm th}$ where $\phi_{\rm th}=3.3 \times 10^{-18} \, \rm TeV^{-1} cm^{-2} s^{-1}$ is the LHAASO KM2A sensitivity threshold at 100 TeV for point-like sources \citep{Celli:2024cny}.
For Cygnus X-1 we instead consider the upper limit of the gamma-ray flux at 100 TeV as derived by the LHAASO collaboration (see Tab.\ref{table sources}).

By comparing the estimated target density with the measured values for a source extension of $R_{\rm s}=50 \, \rm pc $ (assuming similar extension to the detected microquasars), we derive upper limits on $\xi_{\rm Edd}$ for the undetected systems where possible (see the second columns of Table \ref{tab:unresolved_diffusion}).
These limits also translate into upper limits on the accumulated grammage and on the possible contribution of these sources to the CR knee, as summarized in Table \ref{tab:unresolved_diffusion}.

\begin{figure*}[ht]
\centering
\includegraphics[width=0.45\textwidth]{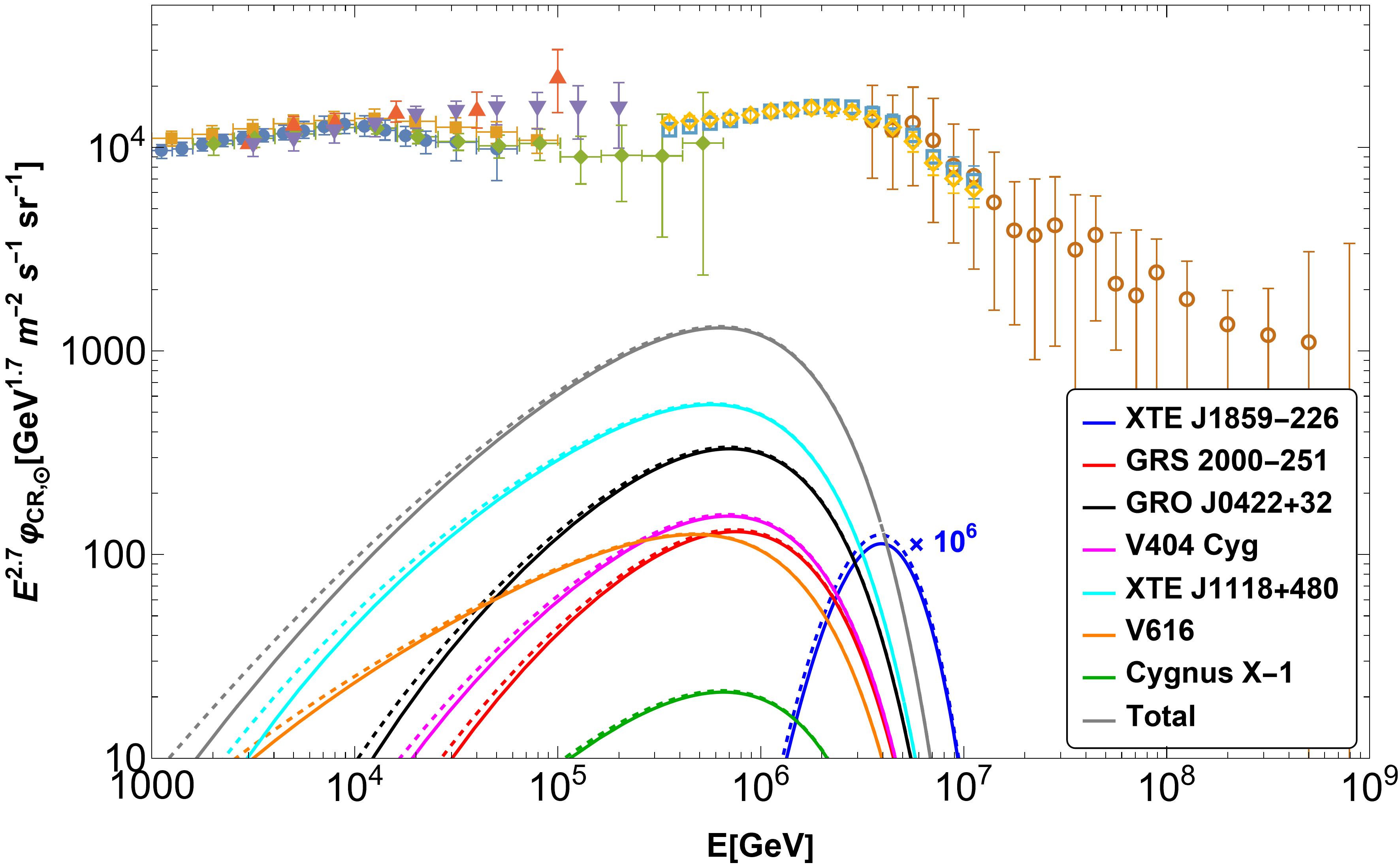}
\includegraphics[width=0.45\textwidth]{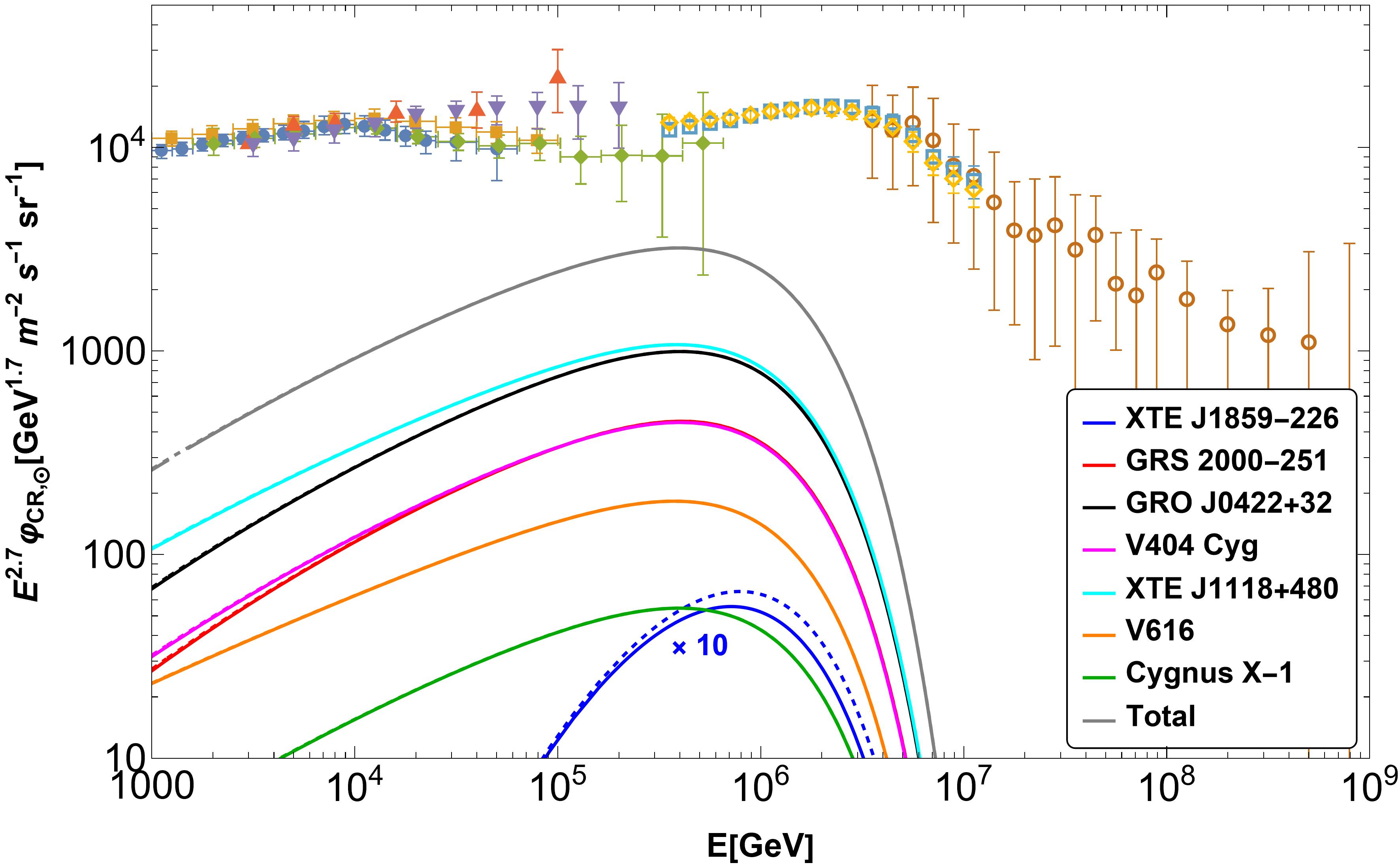}
\caption{Upper limit on the contribution to the CR spectrum for the microquasars undetected for LHAASO. The left (right) panel represent the contribution in case of continuous injection over 100 kyr (1 Myr).
The flux of XTE J1859+226 is scaled by factors of $10^6$ (left panel), and $10$ (right panel), to reduce the dynamic range of the y-axis and improve visual clarity.
}
\label{fig: CR undetected diffuse}
\end{figure*}

Assuming continuous injection over 100 kyr (see left panel of Figure \ref{fig: CR undetected diffuse}), the total contribution to the CR spectrum at 1 PeV remains below $\sim8\%$.
For longer injection timescales, e.g., 1 Myr (see right panel of Figure \ref{fig: CR undetected diffuse}), the contribution can increase up to $17\%$.
It is worth noting that this estimate assumes that the proton spectrum injected by these faint sources is the same as the one inferred for the detected microquasars (power-law with spectral index 2 and cutoff 1  PeV).
Since these sources are Sub-Eddington, they may not be able to accelerate particles up to PeV energies, in which case their contribution to the CR knee would be smaller than the values reported in Table \ref{tab:unresolved_diffusion}.
Another possibility, not explored in this work, is that these sources have recently switched off after accelerating particles up to PeV energies \citep[as proposed in, e.g.,][]{Abaroa:2025ege, Zhang:2026igt}.
In this case, the absence of bright gamma-ray emission would be naturally explained, while the released CRs could produce an even larger contribution to the CR knee.

\begin{table*}[ht]
\caption{Same as Table \ref{tab:diffusion} but for undetected microquasars observed by LHAASO.
All quantities represent upper limits. The gamma-ray flux is assumed to satisfy
$\phi_\gamma(100\,\mathrm{TeV})\le \phi_{\rm th}$ with 
$\phi_{\rm th}=3.3\times10^{-18}\,\mathrm{TeV^{-1}\,cm^{-2}\,s^{-1}}$ (LHAASO KM2A threshold at 100~TeV) with the exception of Cygnus X-1 where instead the upper limit derived from the LHAASO collaboration is used (see Table \ref{table sources}).
Columns report the Eddington ratio $\xi_{\rm Edd}$, grammage $X_{\rm Diff}$ at proton energy of $1$ PeV ($E_{\gamma}\sim100$ TeV), and the contribution $\omega_{\rm Diff}$ to the CR spectrum at 1~PeV for two source ages.
}
\label{tab:unresolved_diffusion}
\centering
\begin{tabular}{lcccc}
\hline\hline
Source &
$\xi_{\rm Edd}(R_{\rm s}=50\,\mathrm{pc})$ &
$X_{\rm Diff}(100\,\mathrm{TeV})$ [$\mathrm{g\,cm^{-2}}$] &
$\omega_{\rm Diff}$ (0.1 Myr) &
$\omega_{\rm Diff}$ (1 Myr) \\
\hline
XTE J1859+226    & 1    & $6.6\times10^{-2}$ & $\ll 1\%$ & $\ll  1\%$ \\
GS 2000+251      & 1   & $4.5\times10^{-3}$ & $0.8\%$ & $2.4\%$ \\
GRO J0422+32     & 0.7    & $1.8\times10^{-3}$ & $2.1\%$ & $5.3\%$ \\
V404 Cyg         & 0.5     & $4.6\times10^{-3}$ & $1\%$ & $2.4\%$ \\
XTE J1118+480    & 1       & $1.4\times10^{-3}$ & $3.3\%$ & $5.7\%$ \\
V616       & $5.9\times10^{-3}$ & $9.6\times10^{-2}$ & $1 \%$ & $0.9\%$ \\
Cygnus X-1 & $2.3\times10^{-2}$ & $6.2\times10^{-3}$ & $0.1\%$ & $0.3\%$  \\
\hline
$\sum$       & $-$ & $0.18<0.4$ & $\sim 8 \%$ & $17.1\%$ \\
\hline
\end{tabular}
\end{table*}

\section{Discussion}
\label{sec: Discussion}
We find that, under standard assumptions for the CR acceleration efficiency, the UHE gamma-rays measured by LHAASO can only be partially explained by $p-p$ interactions. 

The small hadronic contribution inferred for V4641~Sgr is consistent with the conclusions of the HESS collaboration \citep{HESS:2025uyo} and with recent studies \cite[see, e.g.,][]{Wan:2025eoh} suggesting that the emission origin is mainly leptonic.

Concerning the highest energy emission from SS~433, we reached a different conclusion with respect to \cite{LHAASOmicroquasars}, mainly due to the adopted target gas density and the proton luminosity.

The LHAASO collaboration identifies an atomic cloud located at  a distance of $\sim100$ pc from the source, approximated as a spherical cloud with mass $10^{5} \, M_{\odot}$ and radius $40$ pc; corresponding to a target density of $\sim 15 \,\rm cm^{-3}$ within the cloud.
Such a cloud is not clearly visible in the gas maps used in this work.
Even if we adopt the same cloud density and distribute it over a larger spherical region of radius 100 pc, in order to make it compatible with our measurements centered on SS~433, we obtain an average density of $0.96 \, \rm cm^{-3}$.
In the same region, our measurements indicate a gas density about an order of magnitude lower.
In addition, \cite{LHAASOmicroquasars} assume an optimistic proton luminosity above 1 GeV of $10^{39} \, \rm erg \, s^{-1}$, motivated by estimates suggesting that the total luminosity of SS~433 may reach or exceed this value \citep{Marshall:2001ba,Middleton:2018xiq,Fogantini:2022hxv}.
In contrast, in this work we assume a total luminosity above 1 GeV of $3.3\times 10^{39} \, \rm erg \, s^{-1}$, obtained as the product of $L_{\rm Edd} \xi_{\rm Edd}$, and that a fraction $\eta_{\rm CR}=0.1$ is converted into CRs.
This corresponds to a proton luminosity that is only $\sim30 \%$ of the value assumed in \cite{LHAASOmicroquasars}. \\
%
On the other hand, \cite{Bykov:2025xxl} estimate that the power transferred to accelerated protons above 50 TeV in SS433 is $\sim3\times10^{37} \, \rm erg\, s^{-1}$.
This value is in very good agreement with our assumption on the proton luminosity, $\sim3.7\times10^{37} \, \rm erg\, s^{-1}$, obtained assuming the parameters from our benchmark case (see Table \ref{table sources}) and a proton spectrum with $\alpha=2$ and $E_{\rm cut}= 1$ PeV.
With this normalization we find that only $\sim 4\%$ of the gamma-ray signal from SS~433 can be attributed to hadronic interactions.
Our results are therefore compatible with the conclusions of \cite{Bykov:2025xxl} and \cite{HESS:2024rlh}, who suggest that the majority of the very-high-energy gamma-ray emission from SS~433 can be explained within a leptonic scenario.


%
For Cygnus X-3, \cite{LHAASO:2025ysm} report a hint of orbital modulation of the gamma-ray signal above $100$ TeV at the level of $3.2\sigma$, together with a hardening of the gamma-ray signal above $\sim 1$ PeV.
In that work, the emission above $\sim 1$ PeV is interpreted as originating from $p-\gamma$ interactions of freshly accelerated protons with the ultraviolet radiation field of the companion Wolf–Rayet star. 
However, below 1 PeV, the suppression of the gamma-ray flux predicted by this mechanism is too steep to reproduce the observed emission. 
For this reason, \cite{LHAASO:2025ysm} consider the possibility of an additional radiation channel, such as interactions of TeV protons with X-ray photons or hadronic $p-p$ collisions with the ambient gas.

In this work, we find that inelastic $p-p$ collisions can only account for $\sim 2 \%$ of the gamma-ray signal of Cygnus X-3 at 100 TeV. 
However, according to \cite{Veledina:2023zho}, Cygnus X-3 could be up to $10$ times more luminous than assumed in our reference model.
Under this assumption, the hadronic contribution could reach approximately $20\%$ of the observed gamma-ray emission, while the CR contribution to the proton spectrum at Earth would still remain subdominant.
\cite{Wei:2025zkm} proposes an alternative scenario in which the target of the $p-p$ interaction is provided by the stellar wind of the Wolf-Rayet companion star, which is supposedly able to explain the LHAASO observed spectrum of Cygnus X-3.

Another challenge for both the $p-p$ scenario and the $p-\gamma$ interaction with X-ray photons proposed in \cite{LHAASO:2025ysm} is the difficulty in reproducing the observed orbital modulation unless very dense targets are assumed.
In particular, in the $p-p$ case considered in this work, the target density is too low, and the interaction region ($R_{\rm s}=50\,\rm pc$) is too extended to preserve variability on orbital timescales.
Even neglecting the issue related to the size of the interaction region, the $p-p$ interaction timescale at PeV energies is $\tau_{\rm pp}(1 \,{\rm PeV})=(n_{\rm T}\sigma_{\rm pp}c)^{-1}\sim 3\times 10^{14}\,\rm s$, where $n_{\rm T}=2.38 \, \rm cm^{-3}$ is the measured gas around Cygnux X-3 and $\sigma_{\rm pp}\sim 6 \times 10^{-28}\, \rm cm^{2}$ is the total $p-p$ cross-section \citep{Kelner:2006tc}. 
This timescale is many orders of magnitude longer than the orbital period of Cygnus X-3 ($4.8 \,\rm h$). This makes it difficult for such process to account for the observed periodicity of the signal.

Finally, the case in which a source is intermittent is not easily treated within the CR density solutions adopted in this work.
The radial profile of the CR density around the source depends strongly on the duration of active periods $T$, the quiescent intervals $\Delta t$ and the time since the last burst. 
%
In general, intermittency reduces the average CR density around the source, and requires a larger gas density to reproduce the observed gamma-ray flux, which may be incompatible with observations \citep{Evoli:2019wwu}.
%
In this case, microquasars are unlikely to be the dominant contributors to the CR proton spectrum. 
We stress that this conclusion applies only to the population of microquasars observable by LHAASO, as we do not account for the possible existence of additional microquasars outside the sky region covered by the LHAASO detector.
We expect that this would change the conclusion by at most a factor of $2$.
However, if the emission is mainly leptonic or dominated by $p-\gamma$ interactions, the constraint imposed by the gas density is not applicable anymore, and in the latter case the contribution of microquasars to Galactic CRs may still be significant. 

%
In addition, intermittency would further suppress the contribution of these sources to the CR spectrum at Earth by a factor $\sim T/(T+\Delta t)$.
Therefore, the $p-p$ interpretation becomes even more challenging due to a lack of both target material and CRs.


It is also relevant to remark that, in our case, the contribution to the CR proton spectrum is calculated under the assumption of isotropic diffusion in the Galaxy. 
In case of anisotropic CR diffusion, the contribution to the CR proton knee strongly depends on the large-scale Galactic magnetic field.
In particular, when the regular field has a non-negligible component along the perpendicular direction to the Galactic plane, the footprint of each source in the Galactic disc is typically smaller, often resulting in a more clumpy CR distribution \citep{Giacinti:2023upw}.
Moreover, if sources are connected to Earth by regular Galactic magnetic field, their contribution to the CR proton "knee" would be enhanced. On the contrary, if they are not magnetically connected, their contribution could be significantly reduced \citep{Giacinti:2023upw}.
This possibility was investigated in, e.g.,  \cite{Zhang:2026igt}, where it was shown that only Cygnus X-1 and V404 appear connected with the Sun through a magnetic flux tube. 
Clearly this conclusion also strongly depends on the assumed model for the Galactic magnetic field, which introduces significant uncertainty in the interpretation.

\section{Conclusion}
\label{sec: Conclusion}
Recently, the LHAASO collaboration reported the detection of UHE gamma rays from several microquasars, making these systems compelling new candidates for Galactic PeVatrons.

In this work, we model the transport of CRs around the source and in the Galaxy using a two-zone model, in which the diffusion coefficient is significantly suppressed near the source and approaches the Galactic value in the second zone.

Assuming that the gamma-ray emission is produced through hadronic interactions with ambient material, we calculated the grammage required to explain the gamma-ray emission observed by LHAASO at 100 TeV for the considered transport scenario.
We find that the required grammage is compatible with the best-fit value derived from boron-to-carbon observations at a $few$ TeV energies \citep{Evoli:2019wwu}, if CRs are injected continuously over the lifetime of the source, which we assume to be $0.1-1$ Myrs.  This implies that microquasars can potentially contribute significantly to the CR proton spectrum.
However, the grammage requirement could become more challenging if the sources are intermittent rather than continuously active.

Under the assumption of continuous injection, we find that the amount of target material in the vicinity of the sources is generally insufficient to explain the observed gamma-ray flux at 100 TeV. 
%
Increasing the CR acceleration efficiency above the typically used value of 10\%, or adopting smaller diffusion coefficient very close to the Bohm limit, could eventually reconcile the discrepancy for MAXI J1820+070 and GRS 1915+105, while it appears challenging for the remaining sources.
Keeping the acceleration efficiency fixed to 10\%, we derive upper limits on the fraction of the 100 TeV gamma-ray emission that can be produced through $p$–$p$ interactions, finding that only a relatively small fraction of the signal can be explained through this hadronic process.

Our results for SS~433 support the arguments for a leptonic interpretation already proposed by other authors \citep[e.g.][]{HESS:2024rlh, Bykov:2025xxl}.
Our findings also indirectly support the leptonic interpretation proposed for V4641~Sgr in \cite{HESS:2025uyo} and \cite{Wan:2025eoh}.

Finally, we estimated the contribution of these sources to the Galactic CR proton spectrum at Earth assuming injection durations between 0.1 and 1 Myrs and isotropic diffusion in the Galaxy.
We find that the contribution is negligible if the injection lasts $\sim 100$ kyrs and the CR acceleration efficiency is 10\%, but it can become significant, and potentially exceed the observed CR flux, if longer injection times and/or larger efficiency are assumed.
Unfortunately, these estimates strongly depend on the poorly constrained injection histories of these sources and a firm conclusion is difficult to draw.
An additional level of uncertainty is due to the propagation: here we have assumed isotropic diffusion, but anisotropic diffusion along magnetic flux tubes could yield different results from those obtained in this work by enhancing the contribution of sources magnetically connected to us and suppressing that of the others \citep{Zhang:2026igt}.

An additional contribution to Galactic CRs may come also from microquasars not detected by LHAASO. For these systems, we derived upper limits on the Eddington ratio $\xi_{\rm Edd}$ and on the grammage accumulated around the sources.
We have estimated a contribution to the Galactic CR spectrum ranging from $\sim8\%$ for injection lasting $100$ kyr up to $17\%$ for injection lasting $1$ Myr.
This contribution would decrease if these systems operate differently from the detected sources and have a much shorter active phase or a smaller acceleration efficiency. 


\section{Acknowledgements}
\label{sec:acknowledgement}
The Authors are grateful to R. Liu for providing the LHAASO data and for interesting discussions, to K. Koljonen and D. Misra for providing useful information on microquasars, to D. Caprioli, C. Evoli, G. Pagliaroli, F. L. Villante, D.Semikoz and B. T. Zhang for interesting discussions.
The work of V. V. is supported by the European Union’s Horizon Europe research and innovation program under the Marie Skłodowska-Curie grant agreement No. 101208655 (CORNO GRANDE–COnstRaiNing the Origin of Galactic cosmic RAys using gamma-ray and Neutrino Diffuse Emissions).

\begin{appendix}

\section{Microquasar's spectra}
\label{app: spectrum}
In this Appendix, we compare the gamma-ray data of microquasars observed by LHAASO with the hadronic gamma-ray spectra derived in this work, assuming our standard injected proton spectrum: a power law with an exponential cutoff, with spectral index $\alpha = 2$ and cutoff energy $E_{\rm cut} = 1$ PeV.

\begin{figure*}
\centering

\begin{subfigure}{0.48\textwidth}
    \includegraphics[width=\linewidth]{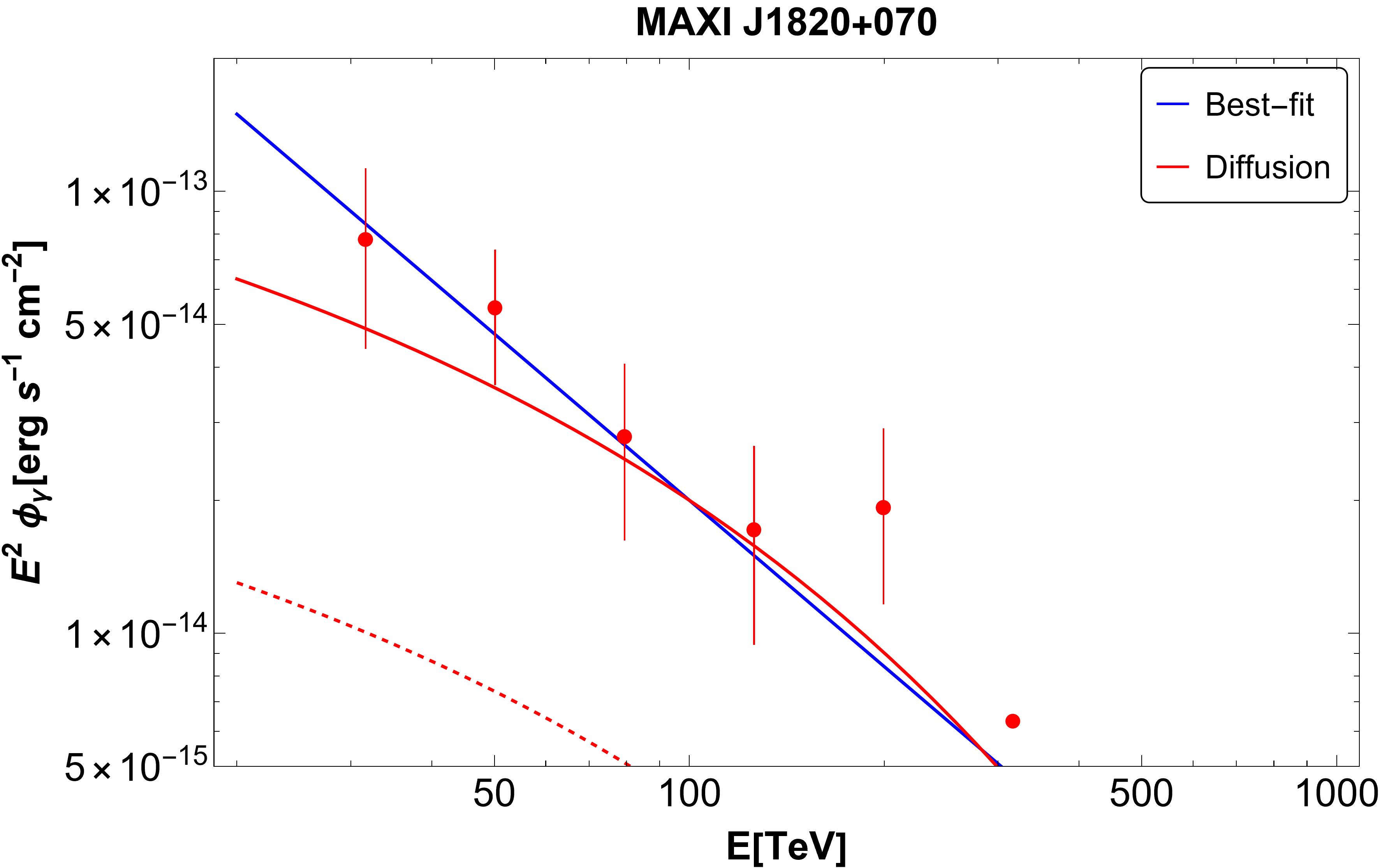}
    \caption{}
\end{subfigure}
\hfill
\begin{subfigure}{0.48\textwidth}
    \includegraphics[width=\linewidth]{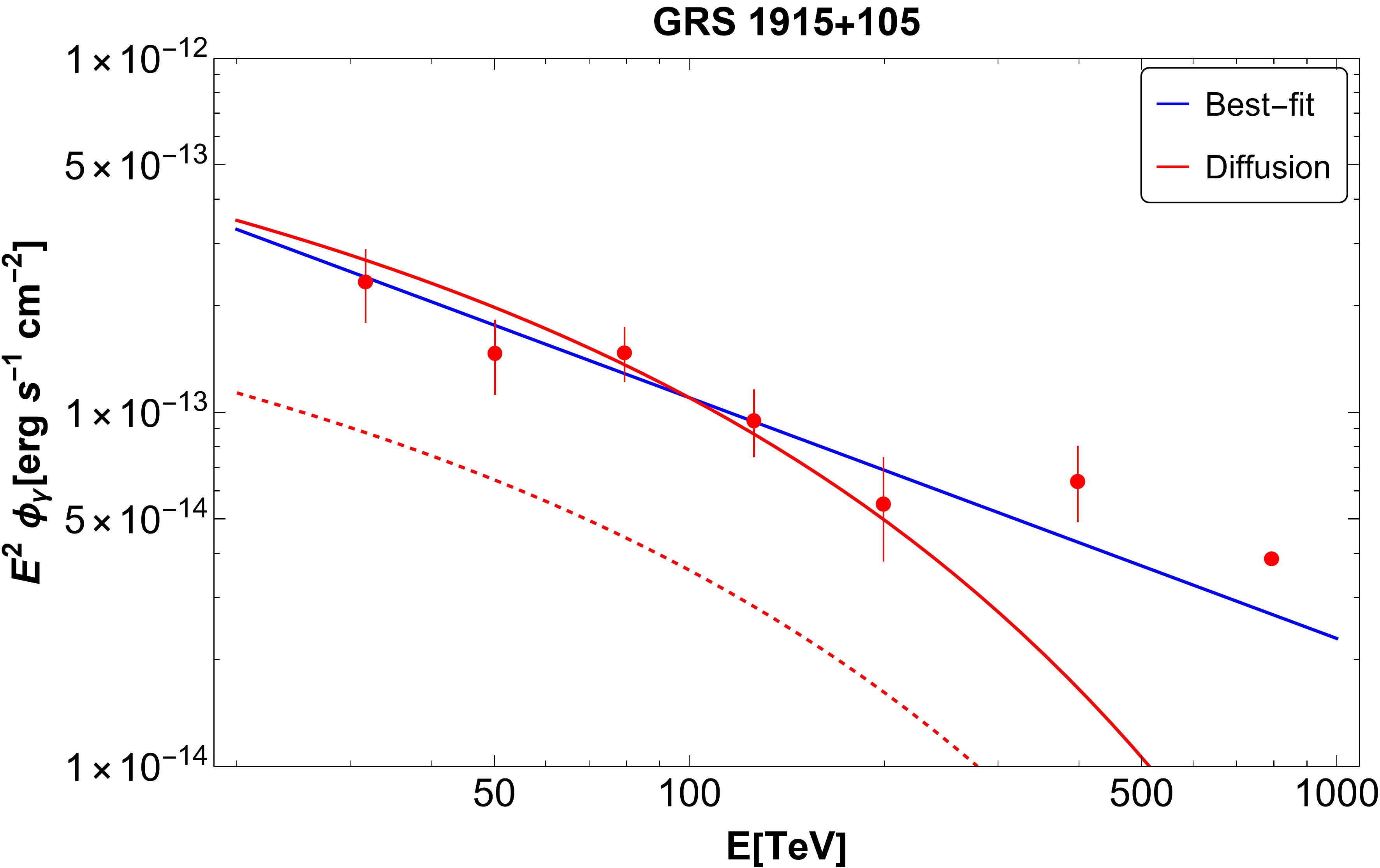}
    \caption{}
\end{subfigure}

\vspace{0.3cm}

\begin{subfigure}{0.48\textwidth}
    \includegraphics[width=\linewidth]{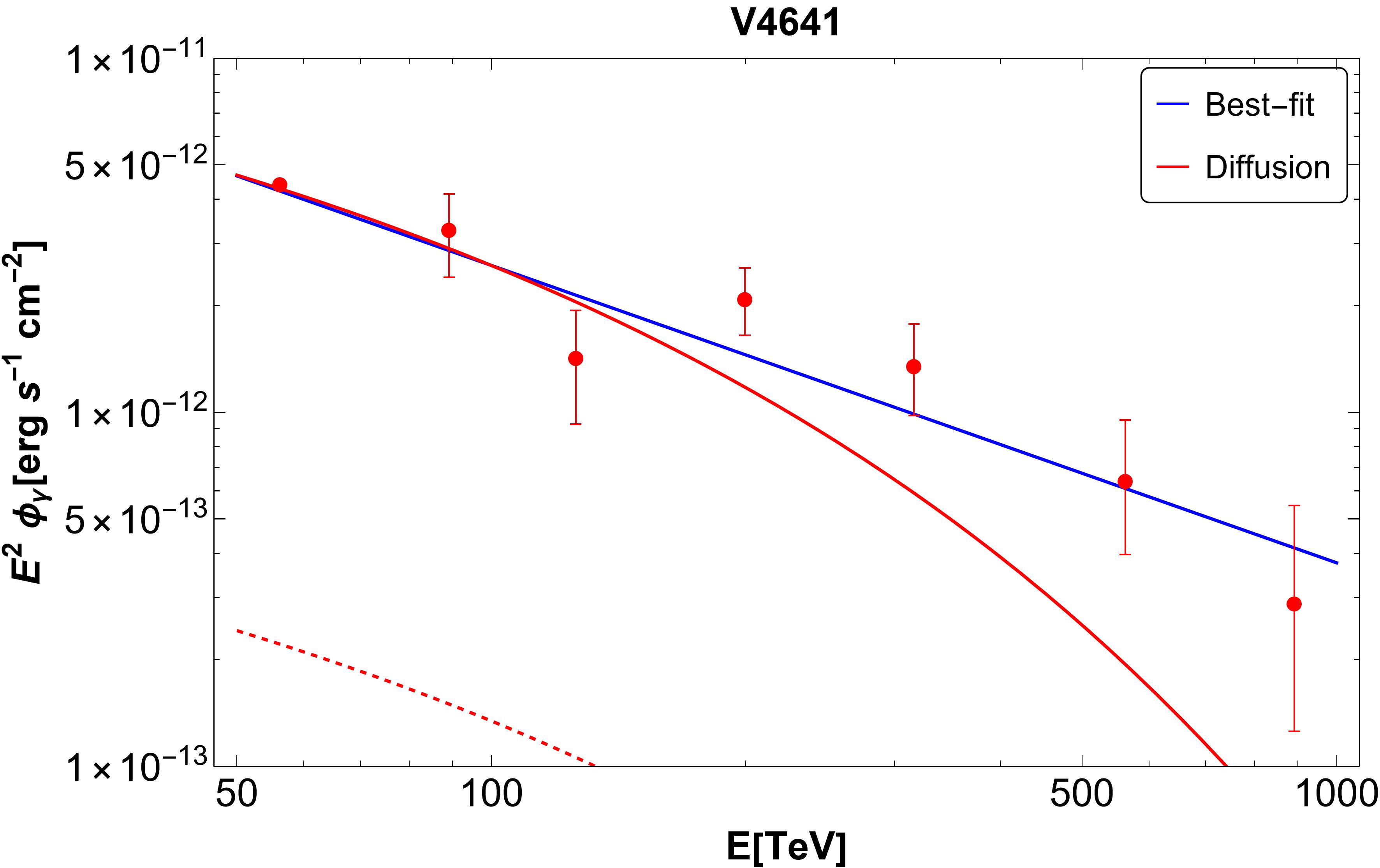}
    \caption{}
\end{subfigure}
\hfill
\begin{subfigure}{0.48\textwidth}
    \includegraphics[width=\linewidth]{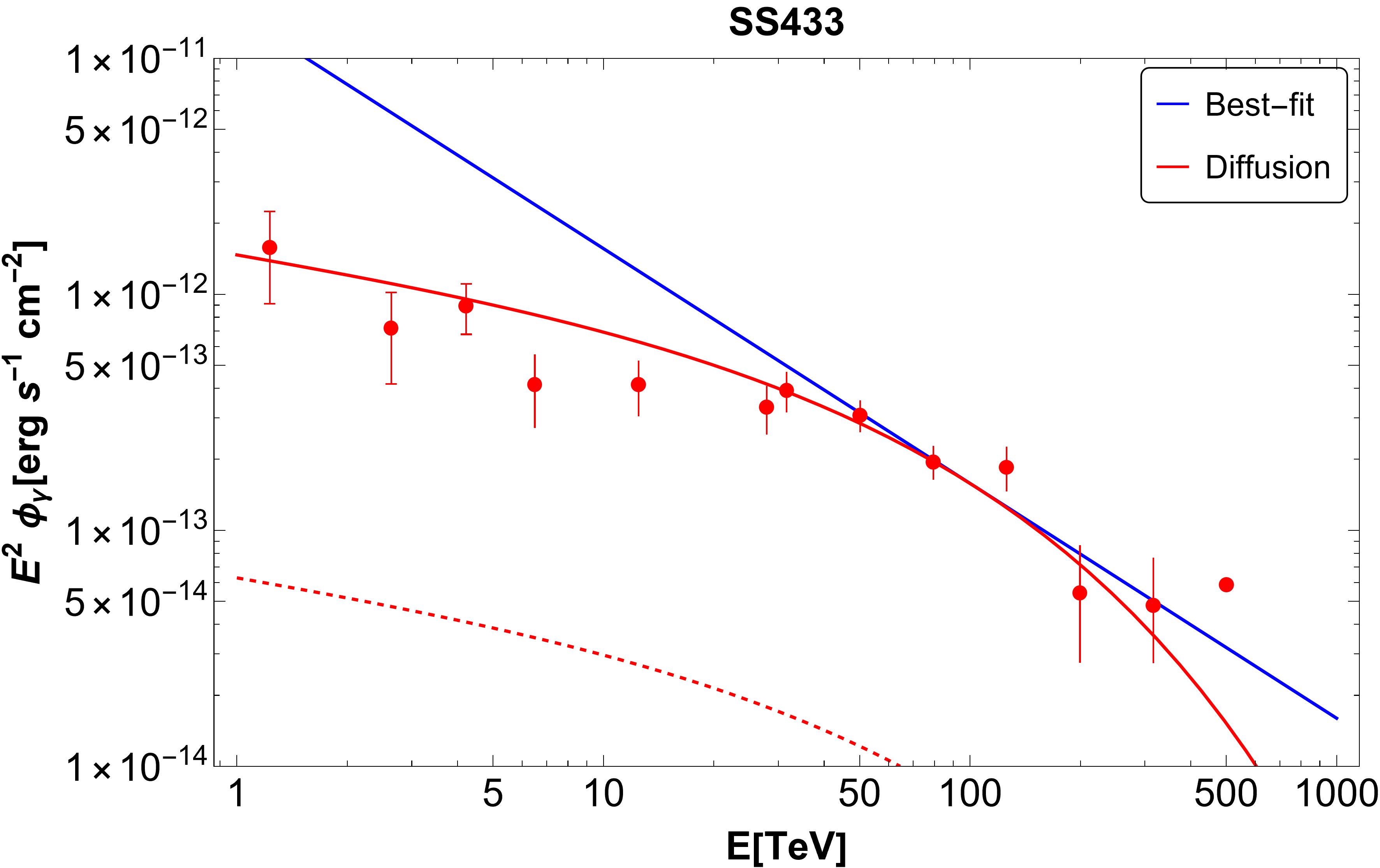}
    \caption{}
\end{subfigure}

\vspace{0.3cm}

\begin{subfigure}{0.48\textwidth}
    \includegraphics[width=\linewidth]{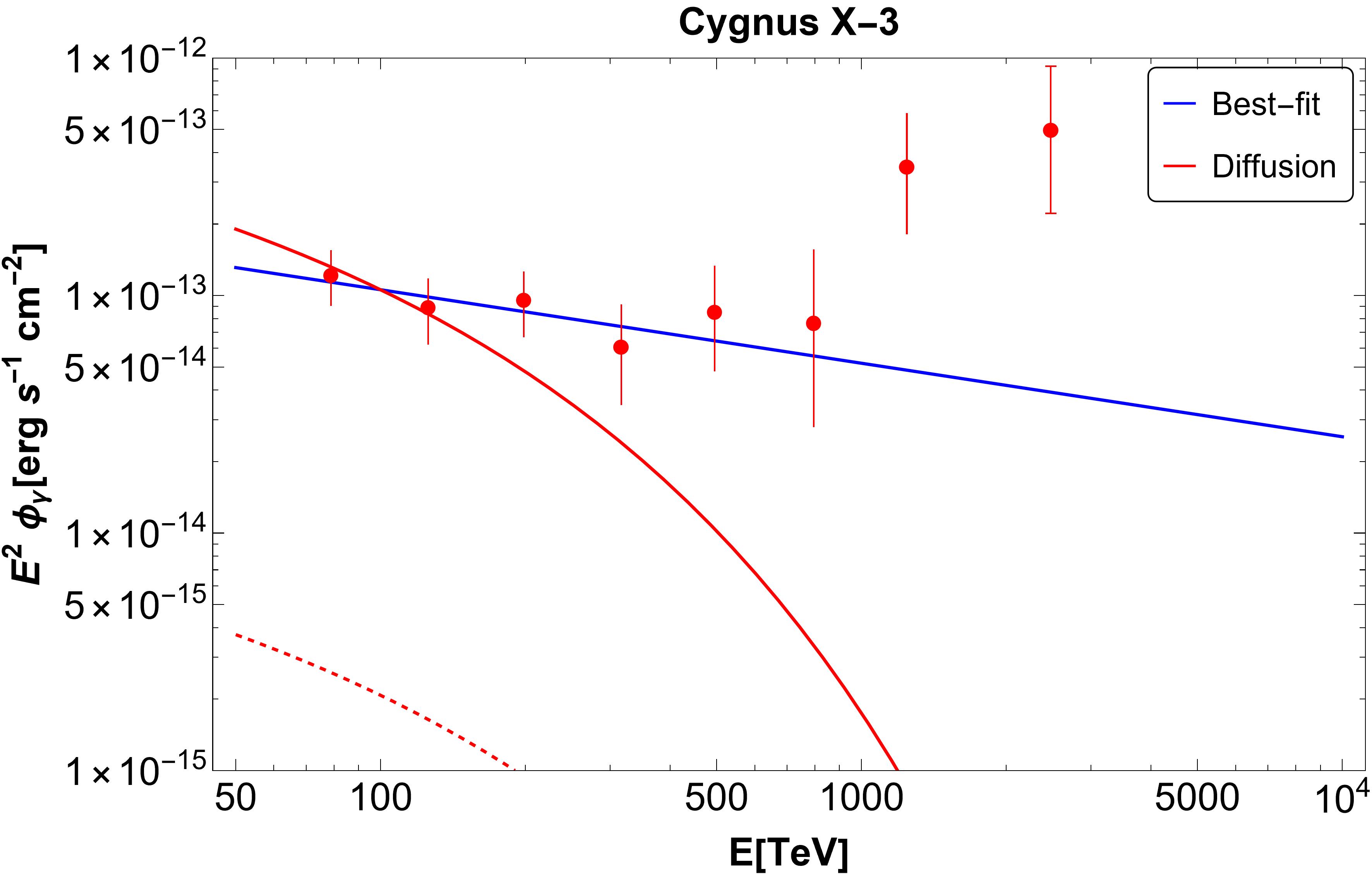}
    \caption{}
\end{subfigure}

\caption{%
Panels (a)–(e) show: the gamma-ray spectra for (a) MAXI~J1820+070, (b) GRS~1915+105, (c) V4641~Sgr, (d) SS~433, and (e) Cygnus X-3. 
The blue lines represent the best-fit power-law spectra. 
The red solid lines show the gamma-ray spectra normalized to the LHAASO data at 100 TeV. 
The red dashed lines are the gamma-ray spectra obtained assuming our standard parameters: CR acceleration efficiency $\eta_{\rm CR} = 0.1$, $r_0=R_{\rm s}= 50$ pc, the measured column density, and the measured $\xi_{\rm Edd}$. 
}
\label{fig: gamma spectra}
\end{figure*}

In Figure \ref{fig: gamma spectra}, the blue lines represent the LHAASO best-fit power-law spectra \citep{LHAASOmicroquasars}. 
For SS~433 and Cygnus X-3, we refitted the data with a pure power law over the energy range relevant to our analysis.
For SS~433, the fit is valid only above 100 TeV. 
For Cygnus X-3, the last two data points are excluded because they are likely associated with $p$--$\gamma$ interactions \citep{LHAASO:2025ysm}, which are not included in our model.
The red solid lines show the gamma-ray spectra predicted by the transport scenario considered in this work and normalized to the LHAASO data at 100 TeV while the red dashed lines show the same spectra obtained using our standard parameters: a CR acceleration efficiency $\eta_{\rm CR} = 0.1$, $r_0=R_{\rm s}=50$ pc, the measured column density, and the measured $\xi_{\rm Edd}$ listed in Table \ref{table sources}.

In general, assuming a higher cutoff energy to 10 PeV enhances the gamma-ray emission produced via $p-p$ at 100 TeV by a factor of $\sim 2.1$ in the transport scenario considered here.

Finally, given that the hadronic contribution is likely subdominant for most of the sources analyzed in this work, the spectral shape of injected hadrons cannot be meaningfully constrained by the current data.

\section{Halo size effect}
\label{app: halo}
For sources located above the Galactic plane and for injection over long periods, the halo size can significantly affect the contribution to the CR spectrum.
In this work, we assume a halo size of $H = 5 \, \rm kpc$ \citep{Evoli:2020szd}.
Since our problem is spherically symmetric, we incorporate the effect of the halo through an energy-dependent correction factor. 
This factor is derived from the solution of the 3D transport equation in Cartesian coordinates, assuming the Earth is located at $x = y = z = 0$ (see, e.g., \cite{Blasi_2012}):
\begin{eqnarray}
n_{\rm H}(E,x_s,y_s,z_s,t_{\rm age})&=&\int_{0}^{t_{age}}dt \frac{1}{\left(4 \pi D_1 K(E)t \right)^{3/2}} 
\nonumber\\
&&\exp{\left(-\frac{x_s^2+y_s^2}{4 D_1 K(E)t}\right)}  \\
\nonumber
&&\sum_{n=-\infty}^{+\infty}(-1)^{n} \exp{\left(-\frac{z_{n}^2}{4 D_1 K(E)t}\right)}
\end{eqnarray}
with $z_n=(-1)^nz_s+2nH$, where $(x_s,y_s,z_s)$ are the coordinates of the source.
We then define the energy-dependent correction factor as:
\begin{equation}
Ratio(E,x_s,y_s,z_s,t_{\rm age})=\frac{n_{\rm H}(E,x_s,y_s,z_s,t_{\rm age})}{n(E,x_s,y_s,z_s,t_{\rm age})}
\end{equation}
where $n(E,x_s,y_s,z_s,t_{\rm age})$ is the CR density in the absence of halo boundaries:
\begin{eqnarray}
n(E,x_s,y_s,z_s,t_{\rm age})&=&\int_{0}^{t_{age}}dt \frac{1}{\left(4 \pi D_1 K(E)t\right)^{3/2}} \\ 
\nonumber
&&\exp{\left(-\frac{x_s^2+y_s^2+z_s^2}{4 D_1 K(E)t}\right)} 
\end{eqnarray}

\section{Stationary vs numerical solution}
\label{app: stationary vs solution}
In this appendix, we compare the stationary solution described in Eq.~\eqref{eq: stationary} with the numerical solution of Eq.~\eqref{eq: transport} and the approximated solution of Eq.~\eqref{eq: approx solution} for an age of $t_{\rm age}=100$ kyr and for $r_0=50, \, 100 $ pc.

\begin{figure}
\includegraphics[width=0.5\textwidth]{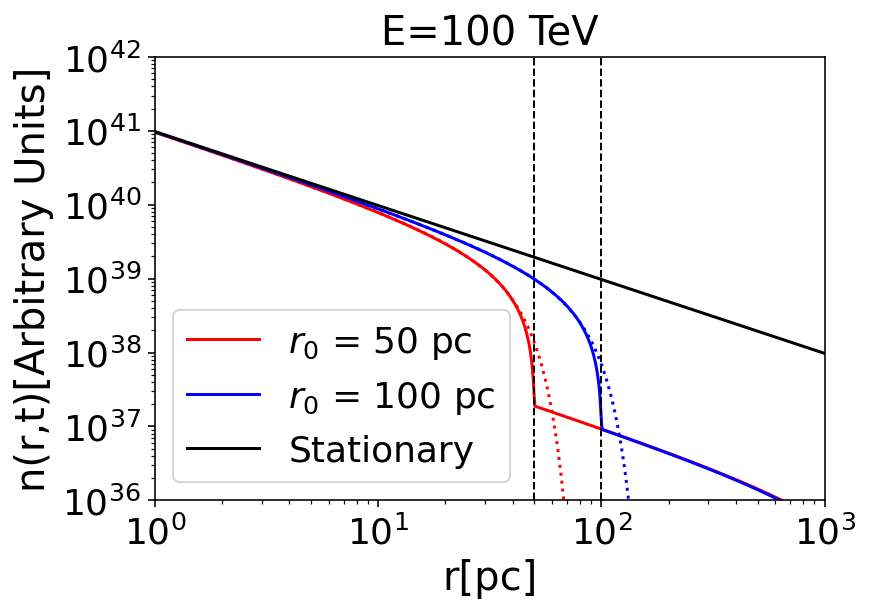}
\caption{ Comparison between stationary, numerical and approximated solutions of the transport equation as a function of distance from the source at fixed energy $E=100$ TeV and for continuous injection over $100$ kyr. The stationary solution is shown as a black line, while the numerical (approximate) solutions for $r_0=50, \, 100 $ pc are shown as solid (dotted) red and blue lines, respectively.}
\label{fig: comparisonPeV}
\end{figure}

In Fig.~\ref{fig: comparisonPeV}, we show the different solutions as a function of radius for particles with energy 100 TeV; the behavior is similar at higher energies.
%
The numerical solution of the transport equation is systematically below the stationary solution.
The gamma-ray flux calculated in Eq.~\eqref{eq: gamma1}, is proportional to the integral over volume of the CR number density. 
Using the stationary solution instead of the numerical one to compute the gamma-ray flux over a region $R_{\rm s}= r_0$ pc leads to an overestimation by factors of $\sim 2.9$ (and this is approximately energy-independent above $\sim10$ TeV).


\end{appendix}

\end{document}